\documentclass[twocolumn]{aastex63}

\accepted{\today}
\submitjournal{ApJ}

\usepackage{comment}
\usepackage{makecell}
\usepackage{amsmath}
\usepackage{lineno}
\usepackage{longtable}
\usepackage{appendix}
\usepackage{xcolor}
\usepackage{url}

\begin{document}

\author[0000-0002-8770-6764]{R\'eka K\"onyves-T\'oth}
\affiliation{Konkoly Observatory, HUN-REN Research Center for Astronomy and Earth Sciences, Konkoly
Th. M. út 15-17., Budapest, 1121 Hungary; MTA Center of Excellence}
\affiliation{Department of Experimental Physics, Institute of Physics, University of Szeged, D\'om t\'er 9, Szeged, 6720 Hungary}

\shorttitle{light curve modeling of SLSNe-I}
\shortauthors{K\"onyves-T\'oth, R}

\correspondingauthor{R\'eka K\"onyves-T\'oth}
\email{konyvestoth.reka@csfk.org}

\graphicspath{{./}{}}

\title{The bolometric light curve modeling of 98 Type I superluminous supernovae using the magnetar- and the circumstellar interaction models reveals surprisingly high ejecta masses}

\begin{abstract}

We present the bolometric light curve modeling of 98 hydrogen-poor superluminous supernovae (SLSNe-I) using three types of power inputs: the magnetar model and two kinds of circumstellar interaction models, applying the constant density and the steady wind scenario. The quasi-bolometric luminosities of the objects were calculated from the ZTF g- and r-band data using the methodology of \citet{chen23b}, and then they were modeled with the Minim code. It was found that the light curves of 45 SLSNe-I can be fitted equally well with both the magnetar and the CSM models, 14 objects prefer the magnetar model and 39 SLSNe-I favor the CSM model.
The magnetar modeling yielded a mean spin period of $P~=~4.1 \pm 0.20$ ms and a magnetic field of $B~=~5.65 \pm 0.43 \cdot 10^{14}$ G, consistently with the literature. However, the ejected mass was estimated to be significantly larger compared to previous studies presenting either multi-color light curve modeling with MOSFiT or bolometric light curve modeling: we obtained a mean value and standard error of 34.26 and 4.67 $M_\odot$, respectively. The circumstellar interaction models resulted in even larger ejecta masses  with a mean and standard error of 116.82 and 5.97 $M_\odot$ for the constant density model, and 105.99 and 4.50 $M_\odot$ for the steady wind model. Although the ejected mass depends strongly on the electron scattering opacity (assumed to be $\kappa~=~$0.2 in this work) and the ejecta velocity, which were estimated to be globally larger compared to earlier studies, our results suggest that SLSNe-I are indeed the explosions of the most massive stars.

\end{abstract}


\section{Introduction}\label{sec:intro}

Superluminous supernovae (SLSNe), the brightest-ever stellar explosions are known for $\sim$2 decades. At first, they were distinguished from traditional supernovae by their extraordinary luminosities, which exceed orders of magnitude the brightness of "normal" stellar explosions \citep[e.g.][]{2007ApJ...668L..99Q,2007ApJ...659L..13O,2007ApJ...671L..17S,2009Natur.462..624G,2011Natur.474..487Q}, therefore a magnitude cut of $-21$ magnitudes was applied to separate them from traditional supernovae \citep[see e.g.][]{2011Natur.474..487Q,2012Sci...337..927G,2015MNRAS.452.3869N,2019ARA&A..57..305G,2021A&G....62.5.34N}.
Later it was found that some slightly fainter objects also resemble spectroscopically to SLSNe,  therefore currently SLSNe are classified by their unique spectra, together with their bright light curves (LCs) \citep[see e.g.][]{2018ApJ...860..100D,2018ApJ...855....2Q,2019ARA&A..57..305G}.
Another difference between normal and superluminous supernovae is the explosion environment: while traditional SNe occur in all types of galaxies, SLSNe prefer low metallicity dwarf galaxies  with high star formation rates (SFRs) \citep[e.g.][]{2013ApJ...763L..28C, 2013sptz.prop10056L, 2014AAS...22412113L, 2015MNRAS.449..917L, 2016MNRAS.458...84A, 2016A&A...593A.115J, 2016ApJ...830...13P,2017ApJ...849L...4C, 2018MNRAS.473.1258S,2018ApJ...857...72H,2021A&G....62.5.34N}.

By their spectra, SLSNe can be divided into two subgroups, similarly to normal supernovae: the members of Type I group have no hydrogen lines in their spectra, while the spectra of Type II SLSNe are rich in hydrogen. The physical difference between these two types is caused by the H envelope of the progenitor star is being preserved until the moment of the explosion, or being stripped by interaction with the companion star/stellar winds.

Several explosion models have been proposed in the past to explain the brightness of SLSNe, including the spin-down of a rapidly rotating neutron star, a so-called magnetar \citep[see][]{2010ApJ...717..245K,2010ApJ...719L.204W}; the vivid interaction with the surrounding circumstellar material \citep[see e.g.][]{2011ApJ...729L...6C,2013AAS...22123305C,2018ApJ...858..115A, 2019RAA....19...63W}, or the pair instability explosion of the most massive stars, with a larger mass than 140 $M_\odot$ \citep{2011ApJ...734..102K, 2017ApJ...836..244W,2015MNRAS.454.4357K,2024A&A...683A.223S}. Initially, pure radioactive decay was used to model the observed luminosities of SLSNe, however, later it was disfavored in most cases, because of the unphysically large nickel mass it required.

Currently, the two most popular models, competing each other are the magnetar scenario and the CSM scenario: there are numerous studies that examine the multi-color, or the bolometric LCs of a specific SLSN, or a small sample of SLSNe, and the conclusion is that in most cases both models can reproduce the observed brightness equally well, but the resulting ejected mass is different \citep[see eg.][]{2014MNRAS.441..289B, 2015AstL...41...95B, 2015MNRAS.449.1215P, 2015ApJ...807L..18N, 2015MNRAS.452.1567C, 2015A&A...584L...5M, 2016Sci...351..257D, 2016ApJ...818L...8S, 2016ApJ...826...39N, 2016ApJ...828...87W, 2016ApJ...831..144L, 2017ApJ...835L...8N, 2017ApJ...835..266T, 2017ApJ...839L...6S, 2017A&A...602A...9C, 2017MNRAS.469.1246K, 2017ApJ...845L...2T, 2017ApJ...845L...8N, 2017MNRAS.470..197Y, 2017ApJ...851L..14W, 2018A&A...610L..10D, 2018MNRAS.475.1046I, 2018ApJ...865....9B, 2018ApJ...866L..24N, 2018A&A...620A..67A, 2018ApJ...868L..32B, 2019ApJ...876L..10E, 2019ApJ...877...20W, 2019MNRAS.488.3783B, 2020ApJ...891...98L, 2020MNRAS.497..318L, 2020ApJ...901...61L, 2021MNRAS.502.2120F, 2021ApJ...921...64B, 2022NewA...9701889K, 2023A&A...670A...7W, 2023MNRAS.521.5418P}. 
This remains true in the studies that examine a larger sample of SLSNe-I
\citep[see e.g.][]{2015MNRAS.452.3869N, 2017ApJ...840...12Y, 2017MNRAS.470.3566C, 2018ApJ...860..100D, 2018ApJ...864...45M, 2018MNRAS.479.4984C, 2019ApJ...874...68C, 2019MNRAS.487.2215A, 2020ApJ...897..114B,2021ApJ...921..180H, 2022ApJ...937...13H, chen23b, chen23a}. In general, the magnetar models tend to show lower ejecta masses compared to the CSM models.

The first semi analytic magnetar and CSM model is proposed by \citet{2012ApJ...746..121C}, who created Minim to fit the bolometric light curves of all types of supernovae, and later tested its validity on multiple objects \citep[see e.g.][]{2013ApJ...773...76C}. Here, in the specific case of SN~2006oz, both the CSM and the magnetar model gave good fit to the observed brightness, while the other examined SLSNe were compatible only with the CSM model. \citet{2014MNRAS.437..703M} revealed that X-ray data helps in distinguishing between the magnetar and the CSM model. Later, the MOSFiT code \citep{2017ascl.soft10006G}, which fits multi-color data emerged, and became the most popular and most widely used code to fit the LCs of SLSNe \citep[see e.g.][]{2017ApJ...850...55N,2018ApJ...869..166V,2021ApJ...921..180H,2022ApJ...937...13H, chen23b}.

It is important to note that it has always been challenging to fit the commonly present light curve undulations of SLSNe using the magnetar model, and in the most bumpy cases, CSM model has been favored. Recently, however, several papers have been published proposing new magnetar models, that can reproduce the light curve bumps of SLSNe, assuming e.g. binary interaction \citep[see][]{2023ApJ...951...61D,2020SciA....6.2732R, 2024MNRAS.527.6455O, 2024arXiv240501224Z}.

In this paper, we present the bolometric light curve modeling of 98 SLSNe-I, which is the largest sample up to date. Our main goal is to test whether the bolometric light curve modeling of SLSNe-I using the Minim code is consistent with the MOSFiT multi-color light curve modeling or the bolometric light curve modeling of a large sample. Up to date, \citet{chen23b,chen23a} have
carried out the analysis of the largest sample, consisting of 77 SLSNe-I from the ZTF survey. They used the MOSFiT code to model the light curves of the objects using both the CSM and the magnetar model. Here, we carry out a comparative analysis with \citet{chen23b,chen23a}, and model the largest dataset of bolometric light curves using the magnetar and the CSM model of Minim code so far. In Section \ref{sec:data} we show the sample selection process and the construction of the bolometric light curves. Section \ref{sec:velocity} describes the velocity estimates from spectroscopy, and Section \ref{sec:lcbol} details the bolometric light curve modeling. Finally, our results are compared with other studies in Section \ref{sec:disc}, and our conclusions are summarized in Section \ref{sec:sum}.

\begin{figure*}[h!]
\centering
\includegraphics[width=16cm]{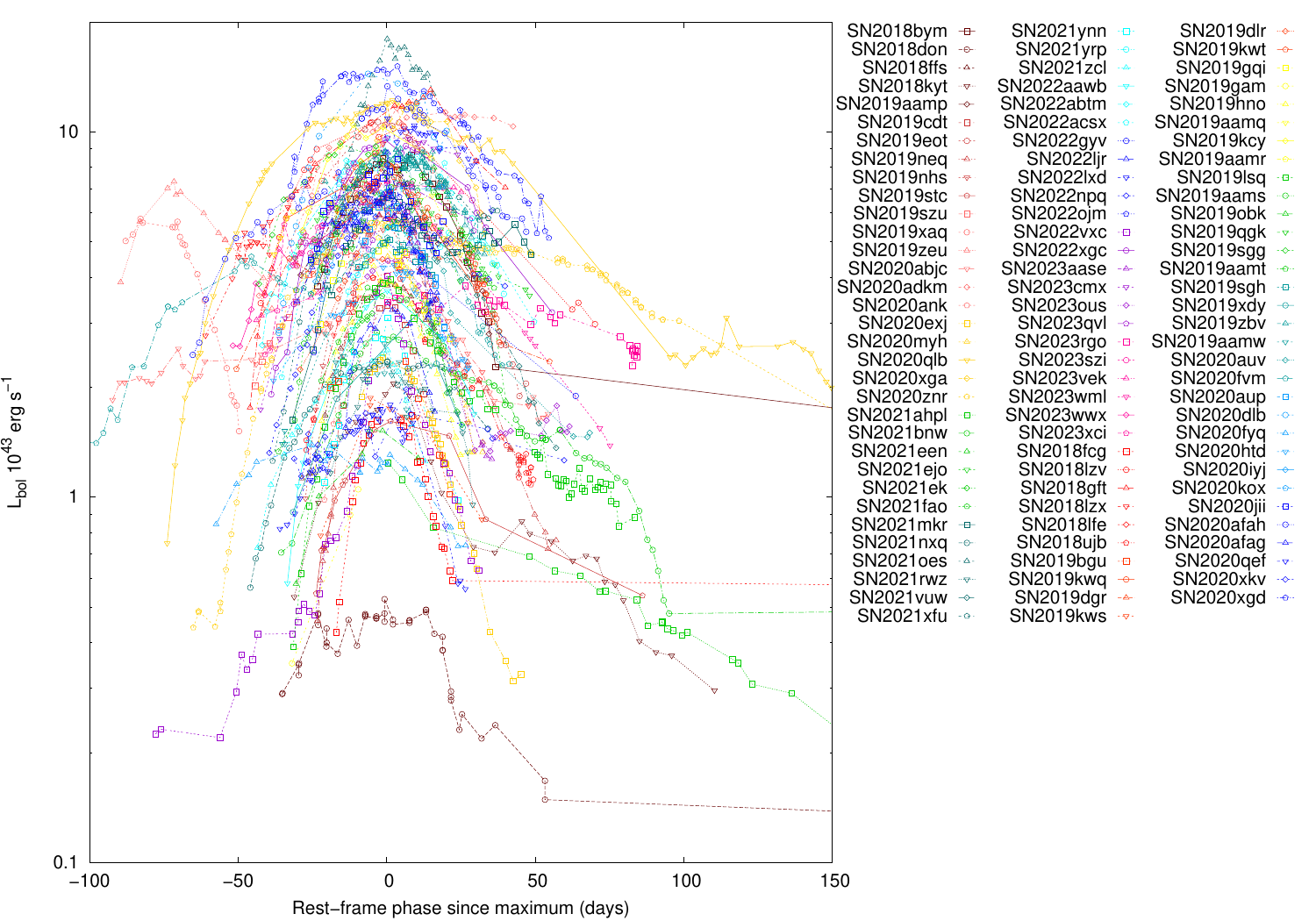}
\caption{Quasi-bolometric light curves of the studied objects plotted with different colors and point styles. }
\label{fig:lbolossz}
\end{figure*}

\section{Light curves of SLSNe-I}\label{sec:data}

\subsection{Sample selection}

A sample of SLSNe-I exploded before 2024 was selected from two publicly available databases. First, WiseRep \citep{2012PASP..124..668Y} was used to find all objects that were classified initially both photometrically and spectroscopically as hydrogen-poor superluminous supernovae. The reason behind using first WiseRep instead of photometric databases was that WiseRep is mostly a spectroscopic database, therefore most objects collected there have published spectra, which are crucial to determine the photospheric velocities of the SLSNe, and thus to calculate reasonable ejecta masses.

Of these 243 SLSNe-I, only 78 had a ZTF name in the WiseRep database, which was then used to obtain g- and r-band light curves of the studied objects. From these 78 objects, further 18 were excluded from the analysis because of having less than 8 points in the light curve and/or having an unrealistic LC for a SLSN (e.g. having a rise time of 7 days, the lack of post-maximum data, or being similar to other types of transients).
The classification of the remaining 60 objects was double-checked using the Transient Name Server\footnote{https://www.wis-tns.org}, and further 4 SNe were excluded from the sample for being classified as other than SLSN-I. The name of the objects removed from the sample, together with the cause of exclusion can be found in Table \ref{tab:removed} in the Appendix.

It is important to note here that our sample selection process differs from that of \citet{chen23b}, which was used as a main comparison study. The two main reasons of choosing a different strategy from \citet{chen23b} was the ambiguity of classification, and the lack of spectroscopic data in the ALeRCE database. From the sample studied by \citet{chen23b,chen23a}, consisting of 78 SLSNe-I exploded before 2021, 60 were not included in our sample at first because of the following reasons: 49 SLSNe-I did not have a ZTF name in WiseRep, and 11 SNe were excluded for being classified as a non-SLSN-I, or having a messy/scarce light curve. However, given that 49 additional objects can significantly broaden the sample, and help to result in a better statistics, it was decided to adopt their data from \citet{chen23b}. Out of these objects, additional 7 was removed, because of insufficiently scarce light-curves.

There are 3 additional SLSNe-I, which did not pass the selection criteria of \citet{chen23b}, but are included in the sample of this paper: SN2018ffs, SN2019xaq and SN2019zeu. According to both the WiseRep and the TNS websites, these objects are SLSNe-I, and their light curves are available in the ALeRCE database as well.

After applying these filters,  $56+42~=~98$ SLSNe-I remained in the sample within the redshift range of 0.06 and 0.553, from which 10 exploded in 2018, 31 in 2019, 22 in 2020, 15 in 2021, 10 in 2022 and 10 in 2023. The basic data, such as the names, the coordinates, the redshift, the interstellar reddening and the date of the maximum are collected in Table \ref{tab:basic}. The redshift (z) of the SLSNe shown in Table \ref{tab:basic} was downloaded from the WiseRep database, the interstellar reddening values (E(B$-$V)) were estimated using the NASA/IPAC Infrared Science Archive \citet{2011ApJ...737..103S} assuming $R_V~=~3.1$, and the time of the maximum in ZTF g-band ($t_{\rm max}$) was obtained from the ALeRCE website\footnote{https://alerce.online} \citep{2021AJ....161..242F}. The approximate $t_{\rm max}$ dates were estimated by eye. More sophisticated methods, such as polynomial fits or Gaussian interpolation were not necessary, since these dates are only reference times fed into the light curve modeling code (see the detailed description of the code in Section \ref{minim}). Thus, these approximate moments of maximum serve only as a zero-point for the time.  The actual moment of explosion is computed with respect to $t_{\rm max}$ by the code as the $t_{\rm ini}$ parameter, and only this latter value has a real physical meaning.

\begin{table*}
\caption{Basic data of the studied SLSNe. The full table is available in the online supplementary material uploaded to Zenodo: \url{ https://doi.org/10.5281/zenodo.14591928}.}
\label{tab:basic}
\begin{center}
\scriptsize
\begin{tabular}{lcccccc}
\hline
\hline
SLSN & Other names$^1$ & R.A.$^1$ & Dec.$^1$ & z$^1$ & E(B$-$V) (mag)$^2$ & $t_{\rm max}$ (MJD)$^3$ \\
\hline
SN2018bym & ZTF18aapgrxo & 18:43:13.424 & +45:12:27.97 & 0.267 & 0.051 & 58274\\
SN2018don & PS18aqo, ZTF18aajqcue & 13:55:08.642 & +58:29:41.98 & 0.0735 & 0.009 & 58280\\
SN2018ffs & ZTF18ablwafp, ATLAS18txu & 20:54:37.160 & +22:04:51.47 & 0.142 & 0.098 & 58370\\
\bf{...} & & & & & & \\
\hline
\end{tabular}
\tablecomments{1: WiseRep; 2:NASA/IPAC Infrared Science Archive; 3: ALeRCE}
\end{center}
\end{table*}

\subsection{Bolometric light curve calculation}\label{lccalc}

To calculate accurate absolute magnitudes and quasi-bolometric bolometric luminosities of the studied objects, light curve corrections are needed to be done for the observed g- and r-band data.

First, the light curves were corrected for Milky Way extiction. Since SLSNe-I tend to occur in low metallicity dwarf galaxies, and because of the lack of host galactic reddening data, corrections for host galactic reddening were neglected. K-correction is crucial as well to calculate precise absolute magnitudes, therefore $K=-2.5 \cdot \rm{log}(1+z)$ was applied to all objects in the sample. According to \citet{chen23a}, this formula of K-correction is not far from the ones estimated using spectroscopic data. Last, the rest-frame light curve phases relative to the moment of the peak brightness were calculated using the $t_{\rm max}$ values presented in Table \ref{tab:basic}.

Afterwards, bolometric light curves were computed using the method of \citet{chen23a}, who applied a bolometric correction related to redshift and $g-r$ color index to the quasi-bolometric luminosities derived using only ZTF g- and r-band data. They tested their correction formula on SLSNe-I having data in all bands of the optical wavelengths, as well as infrared and ultraviolet observations, and found that their calculations can adequately reproduce the bolometric luminosities inferred for more complete SEDs. The equation they used to build the quasi-bolometric light curves give reasonable results for SLSNe-I at low redshifts (having a z in between 0.06 and 0.57), therefore it can be applied for all objects in our sample. The formula is the following:
\begin{equation}
 log (L_{\rm bolo} / L_{\rm gr} )~=~-1.093x^3 + 1.244x^2 - 0.261x + 0.410,   
\label{eq:bolcorr}
\end{equation}
where x refers to the (g$-$r) color index. The further details of the method and its caveats can be found in \citet{chen23b}. The quasi-bolometric light curves of the 60 examined SLSNe-I calculated using Eq. \ref{eq:bolcorr} are plotted together in Figure \ref{fig:lbolossz}.

\section{Velocity estimates}\label{sec:velocity}

\subsection{The method}\label{sec:velmethod}
Before fitting models to the bolometric light curves of the examined SLSNe-I, velocity estimates were carried out. The ejecta velocity is an important parameter, since the ejecta masses of SNe strongly depend on its value. (The ejecta mass calculations are detailed in Section \ref{sec:lcbol} for multiple power inputs of the used modeling code.) Since velocities cannot be constrained from photometric data, spectrum analysis was crucial to give reliable velocity- and ejecta mass estimates.
One of the most common methods to calculate the photospheric velocity of SLSNe is based on the Fe II $\lambda$5169 lines in the spectra \citep[see e.g.][]{2015MNRAS.452.3869N,2017ApJ...845L...8N,chen23b,chen23a}.
However, in some other studies, such as \citet{2020ApJ...900...73K}
it was shown that the identification of the Fe II $\lambda$5169 is ambiguous in a lot of cases, since the spectra of SLSNe usually contain broad and strongly blended features instead of isolated, easily detectable P Cygni profiles. Therefore getting a reliable photospheric velocity value for SLSNe-I is a complicated problem, and requires the modeling of the overall spectra.

Earlier \citet{2021ApJ...909...24K} created a method to estimate the photospheric velocities of SLSNe-I, that combines the spectrum modeling with the cross-correlation process \citep[see e.g.][]{2012MNRAS.419.2783T}.  
To apply this technique, template spectra have to be made, and then cross-correlated with the observed spectra of each object to find the velocity differences ($\Delta v_X$) between the template and the observed spectra. For the cross-correlation, the {\tt fxcor} task in the {\tt onedspec.rv} package of IRAF\footnote{IRAF is distributed by the National Optical Astronomy
Observatories, which are operated by the Association of Universities for Research in Astronomy, Inc., under cooperative agreement with the National Science Foundation. http://iraf.noao.edu} (Image Reduction and Analysis Facility) was used.

In \citet{2021ApJ...909...24K} it was shown that by their pre-maximum spectra, SLSNe-I can be divided into two separate sub-classes: Type W SLSNe-I show a W-shaped absorption blend between 3500 and 5000 \AA, while this well described feature is missing from the spectra of the other group, called Type 15bn SLSNe-I. Therefore, two kind of template spectra were built to be cross-correlated with the pre-maximum spectra. A third kind of template was necessary as well to find the photospheric velocities of the post-maximum spectra. 
\citet{2021ApJ...909...24K} used the SYN++ code \citep{2011PASP..123..237T} to calculate the three types of template spectra, and then a correction formula was created and applied to convert the velocity differences obtained by the fxcor task of IRAF ($\Delta v_X$) to the real velocity differences between the models ($\Delta v_{\rm SN}$). Using this method, one can get reliable photospheric velocity estimates for the spectra of SLSNe-I.

The correction formula between the $\Delta v_X$ and the $\Delta v_{\rm SN}$ velocity differences was found to be the following for Type W SLSNe-I:

\begin{equation}
\Delta v_{\rm{SN}} ~=~ a_0 ~+~ a_1 \cdot \Delta v_{\rm{X}} ~+~ a_2 \cdot \Delta v_{\rm{X}}^2, 
\label{eq:keresztkorr_W}
\end{equation}
, where $a_0 = 155.01~(\pm 82.64)$, $a_1 = 1.68~(\pm 0.03)$ and $a_2 = -2.78 \cdot 10^{-5} ~(\pm 1.63 \cdot 10^{-6})$. 

For Type 15bn SLSNe-I, 

\begin{equation}
\Delta v_{\rm{SN}} ~=~ a_0 ~+~ \sum_{n=1}^{4} a_n \cdot \Delta v_{\rm{X}}^n,
\label{eq:keresztkorr_15bn}    
\end{equation}
was applied, where $a_0= -128.61 ~(\pm 79.92)$, $a_1 = 1.53~(0.06)$, 
$a_2 = 1.09 \cdot 10^{-4}~(3.88 \cdot 10^{-5})$, 
$a_3 =5.45 \cdot 10^{-9}~(7.44 \cdot 10^{-9})$ and
$a_4 = -1.16 \cdot 10^{-12}~(4.27 \cdot 10^{-13})$.

Finally,
\begin{equation}
    \Delta v_{\rm phot} ~=~ a_0 + a_1 \Delta v_{\rm X}
    \label{eq:keresztkorr_maxutan}
\end{equation}
formula was used to calculate the velocities of the post-maximum spectra with $a_0 = 135.62 ~(\pm 54.81)$ and $a1 = 1.51 (0.01)$.

This technique, together with the spectrum templates were adopted from \citet{2021ApJ...909...24K} to obtain reliable velocity estimates for the available spectra. 

\subsection{Application to the sample}\label{sec:applic}

Following the steps described in Section \ref{sec:velmethod}, we inferred the photospheric velocities of the studied SLSNe-I. First, spectra were downloaded from WiseRep, and then corrected for redshift, using the values presented in Table \ref{tab:basic}. Out of the 56 objects selected from WiseRep, 11 had noisy or missing spectra in the database, therefore velocity estimates were done for the remaining 45 SLSNe-I. For the sample adopted from \citet{chen23b}, spectra were available in case of only 9 SLSNe-I. For the velocity calculations, the spectrum nearest to the moment of the maximum was chosen. Table \ref{tab:seb} lists the date of the spectroscopic observations of each SLSNe, together with their rest-frame phase relative to the moment of the maximum light. 

From the template spectra published in \citet{2021ApJ...909...24K}, the following were used in this study: the 10000 km s$^{-1}$ template was applied for the pre-maximum (Type W and Type 15bn) spectra, while the 8000 km s$^{-1}$ template was cross-correlated with the post-maximum spectra. In order to decide which template has to be used in case of a particular object, all 3 templates were cross-correlated with the redshift-corrected observed spectra, and the one that gave the highest correlation was assumed to be the best-fit. Table \ref{tab:seb} shows the name of the best-fit template in each case, the velocity difference estimated by the fxcor task ($\Delta v_X$), and the final velocity value ($v_{\rm SN}$) that was used later in the modeling process.

Figure \ref{fig:fxcor} shows an example of the cross-correlation process on the first member of the sample, SN2018bym. The top left panel shows the redshift versus the value of correlation, while the top right, the bottom left and the bottom right panels plot the observed spectrum (blue) together with the Type W, the Type 15 and the post-maximum template (orange), respectively. 

The resulting velocity values are in the range between 5711 and 30002 km s$^{-1}$, with a mean value of 14707 km s$^{-1}$. 

\begin{table*}
\caption{Velocity estimates for the studied SLSNe-I, using the methodology presented in \citet{2021ApJ...909...24K}. The meaning of the letters in the fifth column are the following: W means that the Type W template gave the best fit to the observed spectrum, 15bn refers to the Type 15bn template, while PM denotes to the post-maximum template, respectively. The last column shows the favored model for each object obtained after the light curve modeling (Mag: the magnetar model, CSM: the CSM1 or the CSM2 model, ALL: all models fit the quasi-bolometric light curve equally well). The full table is available in the online supplementary material uploaded to Zenodo: \url{ https://doi.org/10.5281/zenodo.14591928}.}
\label{tab:seb}
\begin{center}
\scriptsize
\begin{tabular}{lccccccc}
\hline
\hline
SLSN & Date & MJD & Phase  & Template & $\Delta v_X$ & $v_{\rm SN}$ & Favored model\\
     &      &     & (days) &          & (km s$^{-1}$) & (km s$^{-1}$)& \\
\hline
SN2018bym & 2018-05-31 & 58269 & -3.9 & W & -1222.85 & 8146.79 & CSM \\
SN2018don & 2018-06-10 & 58279 & -0.9 & W & 7498.0 & 24284 & CSM \\
SN2018ffs & 2018-09-16 & 58377 & 6.1 & PM & -2724.2 & 12258 & ALL \\
\bf{...} & & & & & & & \\
\hline
\end{tabular}
\end{center}
\end{table*}

\section{Bolometric light curve modeling}\label{sec:lcbol}

\subsection{The code}\label{minim}

After constructing the bolometric light curves, they were modeled using the semi-analytic Minim code \citep{2012ApJ...746..121C}, which fits the radiation-diffusion model of \citet{1980ApJ...237..541A,1982ApJ...253..785A}
to the quasi-bolometric light curve of a supernova. Thus, unlike the commonly used Python-code, MOSFiT, which fits multi-color light curves, Minim requires bolometric data. Minim is a Monte-Carlo based C program, that applies the Price algorithm to localize the absolute minimum of the $\chi^2$ in the parameter space within given parameter bounds. The user can choose from a large variety of power inputs, such as the radioactive decay of Ni and Co, the spin-down of a rapidly rotating magnetar and the interaction with the surrounding circumstellar material (CSM). In this work, the latter two was used to fit the light curves of the studied objects, and then to make constraints on the progenitor properties of each event.

\subsection{Magnetar input}\label{magnetar}

The magnetar spin-down model is widely used to model the light curves of SLSNe-I. According to this scenario, the energy input of the magnetar can be inferred as 
\begin{equation}
L_{\rm inp}(t)~=~{{E_p} \over {t_p}} {{l-1} \over {(1+t/t_p)^l}}
\label{eq:magnetar}    
\end{equation}
\citep[see e.g.][]{2010ApJ...717..245K,2010ApJ...719L.204W}. Here, $E_p$ is the rotational energy of the magnetar, $t_p$ is the magnetic field-dependent spin-down timescale and $l=2$ is used for a magnetic dipole.  From $E_p$ and $t_p$, the initial period of the magnetar and its magnetic field can be expressed as

\begin{equation}
P_{10}~=~ (2 \cdot 10^{50} [{\rm erg/s}]/E_p)^{1/2}; \quad B_{14}~=~(1.3 P_{10}^2/t_{\rm p, yr})^{1/2},
\label{eq:p10b14}    
\end{equation}
\citep[see e.g.][]{2010ApJ...717..245K,2010ApJ...719L.204W}, where $P_{10}$ is given in 10 ms units, $B_{14}$ is given in $10^{14}$ G, and $t_p$ is expressed in years. The magnetar model implemented in the Minim code is based on \citet{2010ApJ...717..245K}, who assumed that the angle between the line of sight and the magnetic axis is 45$^\circ$.

According to this model, the energy of the magnetar termalizes within the ejecta, and then it is assumed to diffuse out through the whole ejecta mass before escaping from the photosphere.
Eq. \ref{eq:magnetar} is incorporated into the general solution for the output luminosity of the SN-photosphere as follows:
\begin{eqnarray}
L(t)=\frac{2 E_{p}}{t_{p}} e^{-[\frac{t^{2}}{t_{d}^{2}}+\frac{t_{h}t} {t_{d}^{2}}]} \cdot \nonumber \\
\cdot \int_{0}^{x} e^{[z^{2}+\frac{t_{h}z}{t_{d}}]} \left[\frac{t_{h}}{t_{d}}+z\right] \frac{1}{(1+yz)^{2}}dz,
\end{eqnarray}
where  $t_h$ is the expansion time-scale defined by \citet{1980ApJ...237..541A}, $x = t/t_{d}$, $y = t_{d}/t_{p}$, and $t_{d}$ is an effective diffusion time.

The parameters of the magnetar model, optimized by the Minim code, are the following: 
\begin{itemize}
    \item $t_{\rm ini}$ (days): the time of the explosion with respect to the moment of the maximum light.
    \item $t_h~=~R_0/v_{\rm SN}$ (days): the expansion time-scale defined by \citet{1980ApJ...237..541A}, where $R_0$ is the initial radius of the progenitor, and $v_{\rm SN}$ is the ejecta velocity. The components of the expansion time-scale can be modeled separately, if one has additional (e.g. spectroscopic) information on the objects, that can make constraints on the ejecta velocity, or the initial radius. In the model, $v_{\rm SN}$ is given in 1000 km s$^{-1}$ and the dimension of $R_0$ is $10^{13}$ cm. Therefore, in this paper, instead of $t_h$, its components were fitted separately.
    \item $E_p$ ($10^{51}$ erg): the rotation energy of the magnetar. 
    \item $t_{\rm diff}$ (days): diffusion time-scale. 
    \item $t_p$ (days): the spin-down time-scale of the magnetar.
\end{itemize}

First, this magnetar model was fitted to the quasi-bolometric light curves of the studied objects. In case of 54 SLSNe-I, where the velocity was determined from spectroscopy, the $v_{\rm SN}$ parameter was fixed to the estimated value, while for the remaining SLSNe-I, that had no available spectra, a velocity range in between 8000 and 30000 km s$^{-1}$ was applied.  The result of the fittings can be found in Figure \ref{fig:bol}$-$ in the Appendix, where the inferred quasi-bolometric light curves are shown with black dots, while the best-fit magnetar model is plotted using orange lines. The values of the best-fit parameters are collected in Table \ref{tab:magnetar_par} in the Appendix. From the fitted parameters, $P_{10}$ and $B_{14}$ can be estimated from Eq. \ref{eq:p10b14}, while the ejected mass, as a key parameter of the explosion can be calculated from the equation of \citet{1980ApJ...237..541A}:

\begin{equation}
M_{\rm ej} ~=~ {{\beta c} \over {2 \kappa}} v_{\rm SN} t_{\rm diff}^2, 
\label{eq:mej}    
\end{equation}

where $\beta~=~13.8$ is an integration constant related to the density profile, and $\kappa~=~0.2$ cm$^2$~g$^{-1}$ is used. 

The choice of the value of $\kappa$ is a complicated, but crucial question, since the ejecta mass strongly depends on it. It was shown by \citet[e.g.][]{2014MNRAS.437..703M} that $\kappa~=~0.2$ is reasonable in the case of the magnetar scenario. However, earlier studies do not agree on the $\kappa$ to be used in the models. For example \citet{2015MNRAS.452.3869N} concluded that $\kappa~=~0.1$ gives more reasonable ejecta masses, than $\kappa~=~0.2$, since with the latter value, the ejecta masses would be lower than expected from SLSNe-I. Later, \citet{2017ApJ...850...55N} find a median of 0.15 in their study, while \citet{2018ApJ...869..166V} get 0.135. In this paper we stick to $\kappa~=~0.2$ in case of all used power inputs for several reasons. The first reason is that we assume that the material surrounding the progenitor is fully ionized below the photosphere, and poor in hydrogen. For a fully ionized plasma containing mostly hydrogen and helium, $\kappa \approx 0.2 \cdot (1+X)$, where X is the mass fraction of hydrogen. Thus, $\kappa < 0.2 $ when the material is not fully ionized and/or rich in metals, while it becomes larger than 0.2, if the matter surrounding the star is rich in hydrogen. In the case of hydrogen-poor SLSNe, the environment of the progenitor is typically rich in carbon, oxygen and helium instead of hydrogen, therefore $\kappa~\approx~0.2$ is a reasonable estimate. Here, we assume that the CSM is poor in hydrogen as well (or the hydrogen shell is so far away that it is difficult to detect), since otherwise we would see hydrogen emission lines, thus, a SLSN-II. 

Furthermore, using $\kappa$ as a fitting parameter in the modeling code increases the uncertainty of the mass calculations and the degeneracy of the problem, since $\kappa$ can be only very poorly constrained from the light curve alone, and it strongly correlates with the ejected mass. As a conclusion, the reason behind the choice of $\kappa~=~0.2$ lies in the assumption of hydrogen-poor material around the progenitor star.

The ejected mass, such as the magnetar spin-down and the magnetic field can be found in Table \ref{tab:magnetar_par2} for all SLSNe-I in the sample. 

Table \ref{tab:magnetar_proxi} gives a summary of the fitted and calculated parameters of the magnetar modeling, together with their dimension, the bounds used in the code, and the mean values for the overall sample.

\begin{table}[h!]
\caption{Parameters of Minim referring to the magnetar powering input, together with their dimensions, the used parameter bounds, and the mean values obtained after modeling the quasi-bolometric light curves of the studied SLSNe-I. The uncertainties in the last column are given as the standard error (standard deviation of the mean).}
\label{tab:magnetar_proxi}
\begin{center}
\scriptsize
\begin{tabular}{lccc}
\hline
\hline
Parameter & Unit & Input bounds & Mean \\
\hline
 $t_{\rm ini}$  & days & $[-100:0]$ & -45.06 $\pm$ 2.27 \\
$R_{\rm 0}$ & $10^{13}$ cm & $[0.1:10]$ & 4.43 $\pm$ 0.26 \\
$E_{\rm p}$ & $10^{51}$ erg & $[0.1:100]$ & 2.64 $\pm$ 0.40 \\ 
$t_{\rm diff}$ & days & $[5:300]$ & 78.79 $\pm$ 7.04 \\
 $t_{\rm p}$ & days & $[1:100]$ & 12.45 $\pm$ 1.33 \\
 $v_{\rm SN}$ & 1000 km s$^{-1}$ & $[8:30]$/fixed & 15.01 $\pm$ 0.58 \\
 \hline 
 $P_{10}$ & 10 ms & calculated & 0.41 $\pm$  0.02 \\
 $B_{14}$ & $10^{14}$ G & calculated & 5.65 $\pm$ 0.43 \\ 
 $M_{\rm ej}$ & $M_\odot$ & calculated & 34.26 $\pm$ 4.67 \\
 \hline 
 $\chi^2$ & & & 3.6787 $\pm$ 0.7021\\
\hline
\end{tabular}
\end{center}
\end{table}

\subsection{CSM interaction input}\label{minim_csm}

Besides the magnetar model, the so-called CSM model, which assumes that the extraordinary luminosity of SLSNe originates from shock heating arising from the violent interaction between the exploding star and the surrounding circumstellar material, is widely used to reproduce the observed light curves of these events. The Minim code can also handle this scenario by taking into account the following 3 physical processes \citep{2012ApJ...746..121C}:
\begin{enumerate}
    \item \textbf{Forward and reverse shock luminosity from the interaction of the ejecta and the circumstellar medium:} According to this scenario, the progenitor of the SN is embedded in a shell of circumstellar material with a density profile of $\rho_{\rm CSM}~=~qr^{-s}$. Here, $\rho_{\rm CSM}$ is the density of the CSM, $q$ is a scaling parameter, and the $s$ parameter is the density profile exponent. 
    For example, $s=2$ denotes a steady wind model, i.e. when the CSM was generated by a constant velocity stellar wind. $s=0$ codes a different mass loss history, when the CSM shell is produced by a strong stellar wind, or a violent, eruptive mass loss of the progenitor star. In this paper, two types of CSM models are used: the constant density model having $s=0$ (hereafter, referred to as CSM) and the steady-wind model having $s=2$ (referred to as CSM2 or CSM W). 
    \item \textbf{The interaction of the ejecta and the CSM with diffusion:} after the energy has been transferred, it is emitted by radiation diffusion. Here, an assumption is made, which presumes that the energy diffuses through the entire ejecta. The model ignores the reheating of the photosphere by the shock that already propagated through the circumstellar material.
    \item \textbf{Hybrid model:} it assumes that the source of the extreme luminosity are the radioactive decay of Ni and Co and the interaction with the circumstellar material. In this paper, the Ni-heating has been switched off in order to reduce the number of free parameters in the models.
\end{enumerate}

It is important to note that, unlike the magnetar model, in the CSM scenario the main parameters of the explosion, such as the ejecta mass, are derived from complex, non-linear formulae, making it difficult to compare its parameters with the other models. The detailed description of this model can be found in \citet{2012ApJ...746..121C}. Here, the input luminosity that is transferred to the ejecta from the forward- and reverse shock energies can be calculated by the following equation:
\begin{eqnarray}
L_{inp}(t)=\frac{2 \pi}{(n-s)^{3}} g^{n\frac{5-s}{n-s}} q^{\frac{n-5}{n-s}} (n-3)^{2} (n-5) \beta_{F}^{5-s}\cdot  \nonumber \\ \cdot A^{\frac{5-s}{n-s}}
(t+t_{i})^{\frac{2n+6s-ns-15}{n-s}} \theta(t_{FS,*}-t)+
\nonumber \\
 + 2 \pi (\frac{A g^{n}}{q})^{\frac{5-n}{n-s}} \beta_{R}^{5-n} g^{n} (\frac{3-s}{n-s})^{3} \cdot \nonumber \\ \cdot
(t+t_{i})^{\frac{2n+6s-ns-15}{n-s}} \theta(t_{RS,*}-t),
\end{eqnarray}
where $\beta_{F}$, $\beta_{R}$ and $A$ are constant values depending on the $n$, the power law exponent of the outer ejecta component and $s$, $\theta(t_{FS,*}-t)$, $\theta(t_{RS,*}-t)$ is the Heaviside step function, which
controls the termination of the forward and reverse shock. Furthermore, $t_{FS,*}$ and $t_{RS,*}$ are the termination time scales for
the forward and the reverse shock, and $t_{i} \simeq R_{p}/v_{SN}$ is the initial time of the CSM interaction that sets the initial value for the luminosity produced by shocks.

The input parameters of Minim referring to the CSM and the CSM2 model are:
\begin{itemize}
    \item $t_{\rm ini}$ (days): similarly to the magnetar model, it codes the days between the explosion and the peak brightness.
    \item $t_{\rm h}$ (days): expansion time-scale. Similarly to the magnetar model, the components of $t_{\rm h}$ can be modeled individually, thus $R_0$ and $v_{\rm SN}$ is treated separately in case of the CSM models as well.
    \item $M_{\rm ej}$ ($M_\odot$): ejected mass.
    \item $M_{\rm dot}$ ($M_\odot$/yr): pre-SN wind mass loss rate, which is connected with the velocity of the stellar wind via the so-called wind parameter, that can be inferred as the ratio of the mass loss rate and the wind velocity. 
    \item $M_{\rm Ni}$ ($M_\odot)$: nickel mass. In this paper it is equaled with 0 to estimate the ejecta masses that originate purely from the interaction.
\end{itemize}

Following \citet{2012ApJ...746..121C}, we set the constants of the models as
\begin{itemize}
    \item $x_0~=~0.1$: the fractional radius of the constant density core within the ejecta.
    \item $n~=~12$: the power-law exponent of the density in the outer part of the ejecta.
    \item $\kappa~=~0.2$ (cm$^2$ g$^{-1}$): the electron scattering opacity.
    \item $s~=~0$ (in case of CSM model) and 2 (in case of CSM 2 model): the CSM density power-law exponent. 
    \item $v_{\rm wind}~=~10$ (km s$^{-1}$): the mass-loss wind velocity.
    \item $d~=~0$: the density exponent in the core.
\end{itemize}

It is important to mention that the efficiency of converting kinetic energy into radiation in the CSM model of Minim is 1.

Figure \ref{fig:bol} show the best-fit CSM (green lines) and CSM 2 (blue lines) models found by Minim to reproduce the observed quasi-bolometric light curves, while in Tables \ref{tab:csm_par} and \ref{tab:csm2_par}, the values of the best-fit parameters for each object are shown. 

Table \ref{tab:csm_proxi} summarizes the names of the optimized parameters of the CSM model and the mean of their best-fit values for the sample. Unlike the magnetar model, here the velocities were not fixed, and a broader range was given to the $R_0$ parameter. The reason to do this was the following: the magnetar model contains $R_0/v_{\rm SN}$, therefore fixing one of them is crucial to avoid parameter correlation. However, the complexity grows in the CSM scenario. The CSM model contains $R_0/v_{\rm SN}$ because of the radiation diffusion, and the velocity (from which the ejecta mass depends strongly) is present in itself in the formula of the shock heating. Because of this, the CSM models do not converge using fixed $v_{\rm SN}$, and that is why the above ranges were used in case of both parameters (see Table \ref{tab:csm_proxi}. (In one case, where the calculated velocity was below 8000 km s$^{-1}$, the applied lower limit was changed accordingly to 5500 km s$^{-1}$.)

\begin{table}[h!]
\caption{Summarizing table of the two CSM interaction inputs of Minim and their average. CSM model codes the constant density model, while CSM 2 refers to the steady-wind model.}
\label{tab:csm_proxi}
\begin{center}
\scriptsize
\begin{tabular}{lcccc}
\hline
\hline
Param & Unit & Bounds & Mean CSM & Mean CSM 2 \\
\hline
 $t_{\rm ini}$  & days & $[0: -100]$ & -62.35 $\pm$ 2.21 & -36.40 $\pm$ 2.43 \\ 
$R_{\rm 0}$ & $10^{13}$ cm & $[0.1:100]$ & 42.13 $\pm$ 2.82 & 64.15 $\pm$ 2.77 \\
 $M_{\rm ej}$ & $M_\odot$ & $[1: 200]$ & 116.82 $\pm$ 5.97 & 105.99 $\pm$ 4.50 \\
 $M_{\rm CSM}$ & $M_\odot$ & $[0: 100]$ & 12.06 $\pm$ 1.74 & 13.90 $\pm$ 1.67 \\
$M_{\rm dot}$ & $M_\odot$/yr & $[0.0001: 2]$ & 0.56 $\pm$ 0.06 & 0.64 $\pm$ 0.06 \\
$v_{\rm SN}$ & 1000 km s$^{-1}$ & $[8:30]$ & 24.78 $\pm$ 0.62 & 21.64 $\pm$ 0.69 \\ 
 \hline 
 $\chi^2$ & & & 1.7437 $\pm$ 0.1251 & 1.8829 $\pm$ 0.1923 \\ 
\hline
\end{tabular}
\end{center}
\end{table}

\section{Results and discussion}\label{sec:disc}

\subsection{Favored models}\label{sec:favored}

\begin{figure}[h!]
\centering
\includegraphics[width=8cm]{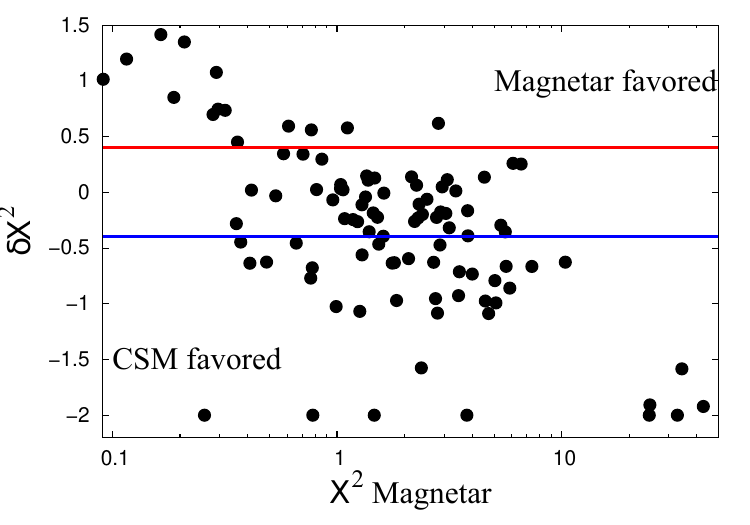}
\caption{ The reduced $\chi^2$ values from the magnetar modeling vs. the difference of the $\chi^2$ between the magnetar and the CSM1/CSM2 models. The region above the red line (i.e. $\delta \chi^2 > 40\%$) show the SLSNe-I that favor the magnetar scenario, while below the blue line, CSM favored objects  (i.e. $\delta \chi^2 < -40\%$) are shown.}
\label{fig:favored}
\end{figure}

To follow the methodology of \citet{chen23a}, we examined which of the applied models fit the quasi-bolometric light curves better. For this we used the reduced $\chi^2$ values derived by Minim. Figure \ref{fig:favored} shows the $\chi^2$ inferred for the magnetar model  against the $\delta\chi^2$ difference between the magnetar model and the CSM model. For the "best-fit" CSM model, we chose either the CSM or the CSM2 model depending on of which $\chi^2$ value was estimated to be the lower. Similarly to \citet{chen23a}, the $\delta\chi^2$ was derived as

\begin{equation}
 \delta \chi^2 = {{(\chi^2_{\rm CSM} - \chi^2_{\rm Magnetar})} \over { 0.5 \cdot (\chi^2_{\rm CSM} + \chi^2_{\rm Magnetar}) }}
\end{equation}\label{eq:deltakhi}
.

We found that 45 SLSNe-I in the sample can be fitted equally well using both the magnetar and the CSM models (according to \citet{chen23a} equally well is defined by  ($|\delta \chi^2| < 40\%$; between the red and the blue lines in Figure \ref{fig:favored}), 14 objects favor the magnetar model ($\delta \chi^2 > 40\%$; above the red line in Figure \ref{fig:favored}) and 39 SLSNe-I favor the CSM model ($\delta \chi^2 < -40\%$; below the blue line in Figure \ref{fig:favored})
In the paper of \citet{chen23a}, 47 examined objects could be fitted equally well with both types of models, 7 SLSNe-I preferred the magnetar model and in the case of 16 SLSNe-I, the CSM model was favored. The last column of Table \ref{tab:seb} shows the preferred model for all studied objects.
Out of the overlapping part of the sample analyzed in this paper and the sample of \citet{chen23a}, SN2018don, SN2018kyt, SN2018cdt, SN2020auv, SN2020dlb, SN2020qef and SN2020afag favored the CSM model consistently. SN2019nhs, SN2019stc, SN2020fvm and SN2020aamw was fitted well using all types of models in this paper, while these two SNe favored the CSM model according to \citet{chen23a}. It was also revealed by \citet{chen23a} that SNe showing light curve undulations usually show better fits with the CSM models, which is true for the present sample as well. The names of the 7 SLSNe-I that prefer the magnetar model in \citet{chen23a} were not mentioned, therefore comparison could not be made.

\subsection{Comparison of the magnetar modeling to the literature}

Multiple studies have been written in the past in which the multi-color or the bolometric light curves of a large sample of SLSNe were modeled, and physical parameters of the explosion were estimated. Most of these papers use the magnetar model, while some of them \citep[see e.g.][]{chen23b} applies the CSM model as well. This section compares the results of the magnetar modeling of several studies, while in Section \ref{sec:chen} we discuss the detailed comparison between our results and the modeling of \citet{chen23b,chen23a}.

The first study about the modeling of a larger sample of SLSNe-I was carried out by \citet{2015MNRAS.452.3869N}, who parameterized the light curve shapes by estimating rise- and decline timescales, looked for correlation between them, and then modeled the bolometric light curves of 24 objects using the magnetar and the CSM models. They found that the magnetar model gives better fits in most cases, and their calculations resulted in ejecta masses ranging from 3 to 30 $M_\odot$ with an average of 10.0  $M_\odot$.

\citet{2017ApJ...840...12Y} searched for relationships between gamma-ray bursts (GRBs) and magnetar-driven supernovae. For this purpose they modeled the bolometric light curves of 31 SLSNe-I using the magnetar model and calculated the physical parameters of the explosion. They revealed that the light curve evolution rate correlates with the magnetic field, and the ejected mass can be related to the rotational energy of the magnetar. Their modeling resulted in the following mean values for the key parameters: $M_{\rm ej, avg}~=~4.0 M_\odot$, $P~=~3.5$ ms and $B~=~2 \cdot 10^{14}$ G. 

In the paper of \citet{2017ApJ...850...55N}, the MOSFiT multi-color light curve modeling of a large sample of SLSNe-I consisting of 31 objects appeared, which code has now became the most commonly used. Their study showed how flexible the magnetar model is, and resulted in the following parameter values: $M_{\rm ej, avg}~=~4.8 M_\odot$, $P~=~2.4$ ms and $B~=~2 \cdot 10^{14}$ G.

Next, \citet{2018ApJ...860..100D} studied 26 SLSNe-I from the Palomar Transient Factory by modeling their light curves using the radioactive decay and the magnetar input. They concluded that 1-10 $M_\odot$ of Ni would be required to reproduce the late-time light curves. Alternatively, they used the magnetar model, which could reasonably fit most of the light curves. They calculated $P~=~2.55$ ms and $B~=~3 \cdot 10^{14}$ G.

The paper of \citet{2018ApJ...864...45M} focused on the X-ray study of 26 low redshift SLSNe-I, and searched for a relationship between SNe, GRB jets and magnetar engines. From their data they gave limits to the explosion parameters as $M_{\rm ej} > 20 M_\odot$,  $B~>~2 \cdot 10^{14}$ G and $M_{\rm dot} < 2\cdot 10^{-5} M_\odot$ / yr.

\citet{2018ApJ...869..166V} continued the work of \citet{2017ApJ...850...55N}, and analyzed a sample of 58 SLSNe-I, from which 20 were newly modeled and 38 were taken from the earlier sample of \citet{2017ApJ...850...55N}. Similarly to the parent article, the authors of this study utilized the MOSFiT code as well to reproduce the observed multi-color light curves of the events. The parameter values they obtained are $M_{\rm ej, avg}~=~6.16 M_\odot$, $P~=~2.6$ ms and $B~=~5.52 \cdot 10^{14}$ G.

\citet{2019MNRAS.487.2215A} studied 21 SLSNe-I from the Dark Energy Survey (DES) by modeling their bolometric light curves using the magnetar model. Their main conclusion was that the magnetar model alone cannot fit properly the observed light curves, and therefore combined models are required to obtain proper fits. Their average values of the fitted parameters are $M_{\rm ej, avg}~=~7.54 M_\odot$, $P~=~6.3$ ms and $B~=~4.26 \cdot 10^{14}$ G.

\begin{table*}
\caption{Parameters of the magnetar from the literature.}
\label{tab:irodalom}
\scriptsize
\begin{center}
\scriptsize
\begin{tabular}{lc|ccc|ccc|ccc|}
\hline
\hline
Paper & Sample size & \multicolumn{3}{|c|}{P (ms)} & \multicolumn{3}{|c|}{$B_{14}$ (10$^{14}$ G)} & \multicolumn{3}{|c|}{$M_{\rm ej} (M_\odot)$} \\
      &            & Avg & Min & Max & Avg & Min & Max & Avg & Min & Max \\
\hline 

Nicholl+15 & 24 & -- & -- & -- & -- & -- & -- & 10.0 & 3 & 30 \\
Yu+17 & 31 & 3.5 & 1.37 & 12.04 & 2 & 1.33 & 11.8 & 4.0 & 0.81 & 17.89 \\
Nicholl+17 & 38 & 2.4 & 1.2 & 4 & 2 & 0.5 & 4.5 & 4.8 & 2.2 & 12.9 \\
DeCia+18 & 14 & 2.55 & 0.06 & 4.4 & 3 & 1.13 & 5.18 & -- & -- & -- \\
Margutti+18 & 26 & -- & -- & -- & -- & 2 & -- & -- & 20 & -- \\
Villar+18 & 58 & 2.6 & 1.11 & 6.85 & 5.52 & 0.15 & 19.52 & 6.16 & 1.09 & 68.8 \\
Angus+19 & 15 & 6.3 & 3.64 & 18.27 & 4.26 & 1.05 & 12.47 & 7.54 & 0.23 & 19.66 \\
Hsu+21 & 21 & 4.97 & 0.79 & 13.61 & 4.18 & 0.08 & 18.33 & 7.56 & 1.54 & 30.32 \\
Hsu+22 & 19 & 4.54 & 1.17 & 10.83 & 6.02 & 0.2 & 19.15 & 10.37 & 1.16 & 47.89 \\
Chen+23 & 77 & 2.64 & 0.93 & 9.05 & 2.45 & 0.18 & 12.3 & 5.03 & 1.17 & 47.92 \\
This paper & 98 & 4.1 & 0.3 & 12.2 & 5.65 & 0.1 & 24.43 & 34.26 & 1.53 & 198.1 \\  
\hline
\end{tabular}
\end{center}
\end{table*}

\citet{2021ApJ...921..180H} studied 21 SLSNe from the DES in order to explore the redshift evolution of the explosion parameters. They used MOSFiT to model the multi-color light curves, and obtained $M_{\rm ej, avg}~=~7.56 M_\odot$, $P~=~4.97$ ms and $B~=~4.18 \cdot 10^{14}$ G. They concluded that there is no redshift dependence in these parameters.

\citet{2022ApJ...937...13H} analyzed 19 photometrically classified SLSNe-I from the Pan-STARRS1 survey in order to test how machine learning based classification works in case of superluminous supernovae. As a side project, they fitted the magnetar model of MOSFiT to the observed light curves, and resulted in  $M_{\rm ej, avg}~=~10.37 M_\odot$, $P~=~4.54$ ms and $B~=~6.02 \cdot 10^{14}$ G.

\citet{chen23b,chen23a} carried out the detailed analysis of 77 SLSNe-I from the Zwicky Transient Facility (ZTF), and modeled their light curves using the magnetar- and the CSM-model of MOSFiT. Their magnetar fitting gave the mean values of  $M_{\rm ej, avg}~=~5.03 M_\odot$, $P~=~2.64$ ms and $B~=~2.45 \cdot 10^{14}$ G. The further comparison with \citet{chen23b}
is discussed in the next section  (Section \ref{sec:chen}).

Table \ref{tab:irodalom} summarizes the P, B and $M_{\rm ej}$ values of the previous studies, showing the number of examined SLSNe, and the mean, minimum and maximum values of the parameters referring to the magnetar model.

\begin{figure}[h!]
\centering
\includegraphics[width=8cm]{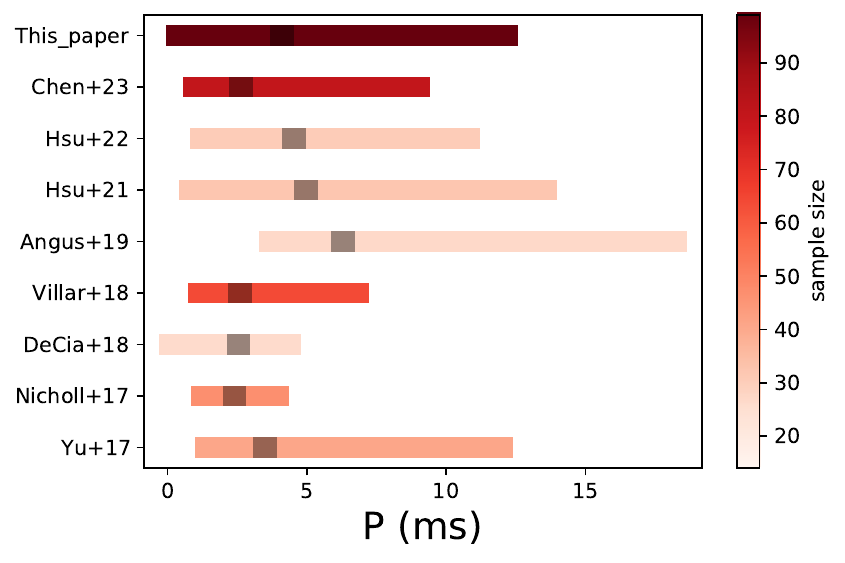}
\includegraphics[width=8cm]{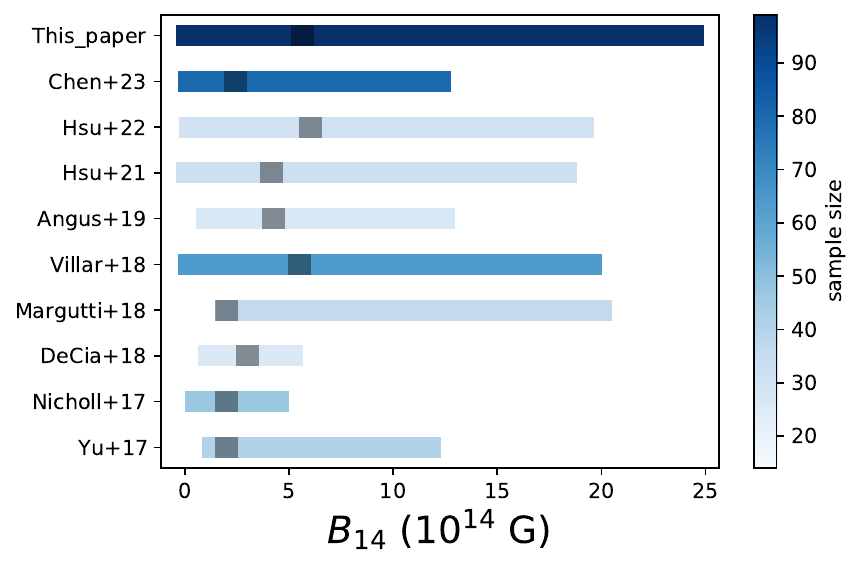}
\includegraphics[width=8cm]{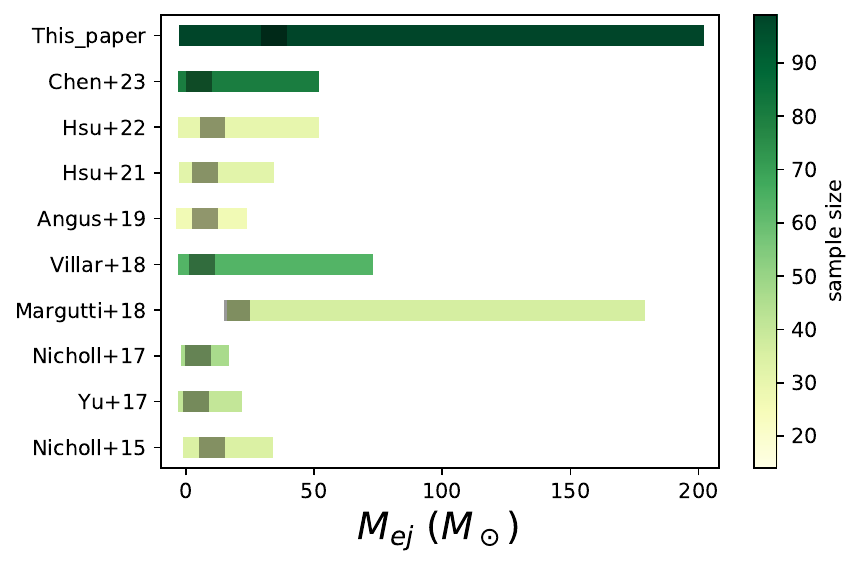}
\caption{Comparison between the P (upper panel, red colors), the B (middle panel, blue colors) and the $M_{\rm ej}$ (bottom panel, green colors) parameters of multiple earlier studies and this paper. The darker the color, the larger the sample. Grey rectangles show the average value of the parameters, while the strips span from the minimum to the maximum of their values.}
\label{fig:irodalom}
\end{figure}

Figure \ref{fig:irodalom} compares these earlier results graphically to the P, B and $M_{\rm ej}$ values calculated in this paper using Minim. The main difference between this paper and other studies is that here the bolometric light curves were modeled, while in most cases (in 7 out of the 10 studies used for comparison) MOSFiT was used to fit multi-color light curves. Figure \ref{fig:irodalom} shows that the P and B values estimated in this study are consistent with others in the literature, while the ejected masses are systematically larger compared to previous studies. The only exception is \citet{2018ApJ...864...45M} who also found higher ejecta masses, although for a smaller sample of SLSNe.

\subsection{Comparison with Chen et al. (2023a,b)}\label{sec:chen}

\subsubsection{The comparison of the magnetar models}

In a way, this study is a continuation of the paper of \citet{chen23b, chen23a}, where ZTF data of SLSNe were used to compile bolometric light curves. In this paper, a test is made to find out if the parameters of multi-color light curve fitting using MOSFiT are consistent with bolometric light curve modeling using Minim.

\begin{figure*}[h!]
\centering
\includegraphics[width=6cm]{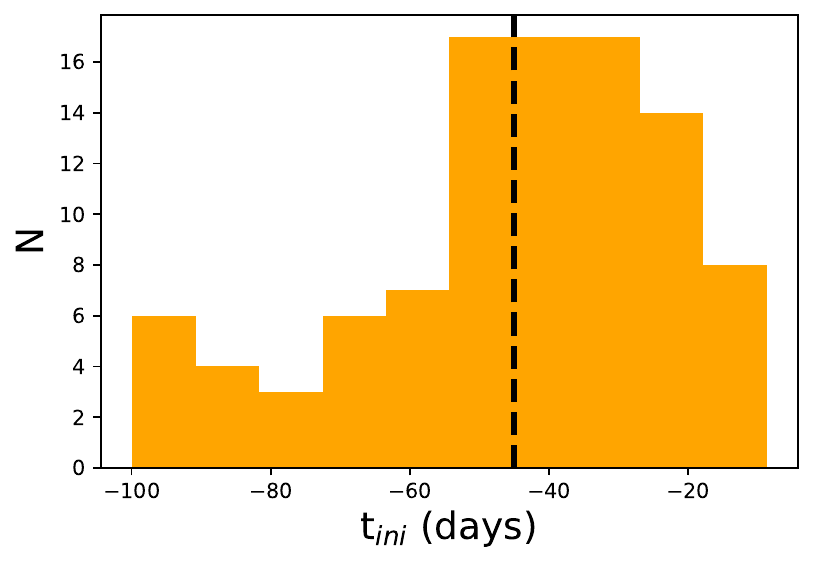}
\includegraphics[width=6cm]{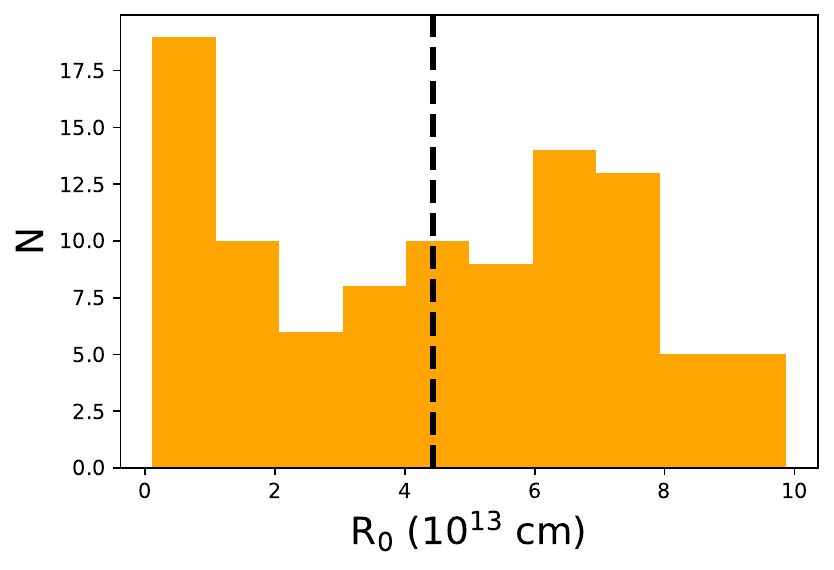}
\includegraphics[width=6cm]{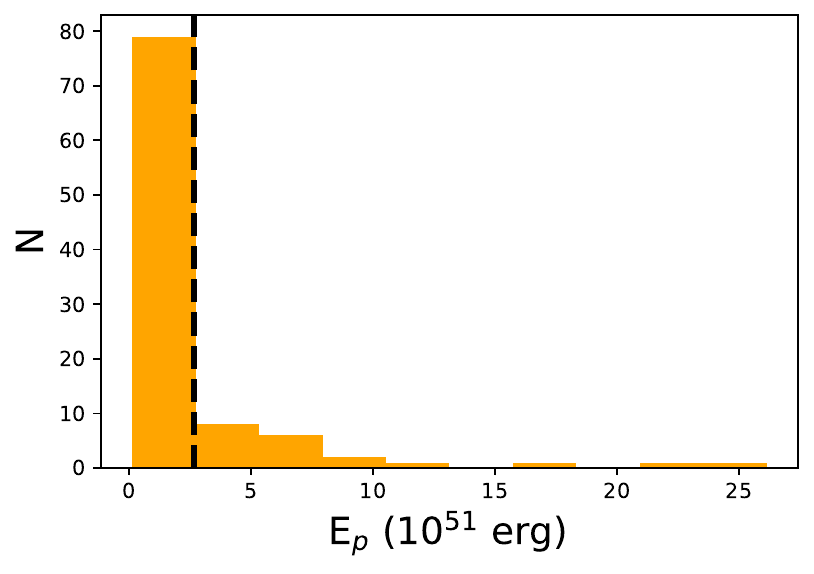}
\includegraphics[width=6cm]{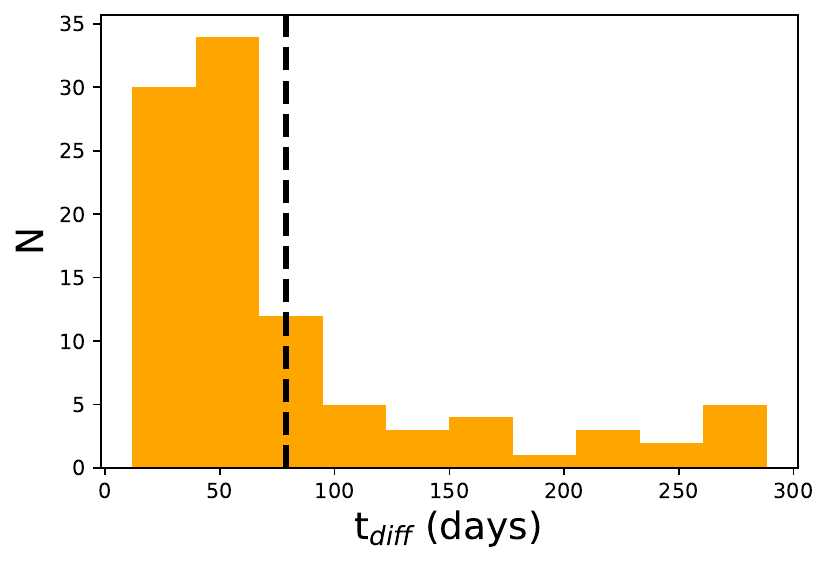}
\includegraphics[width=6cm]{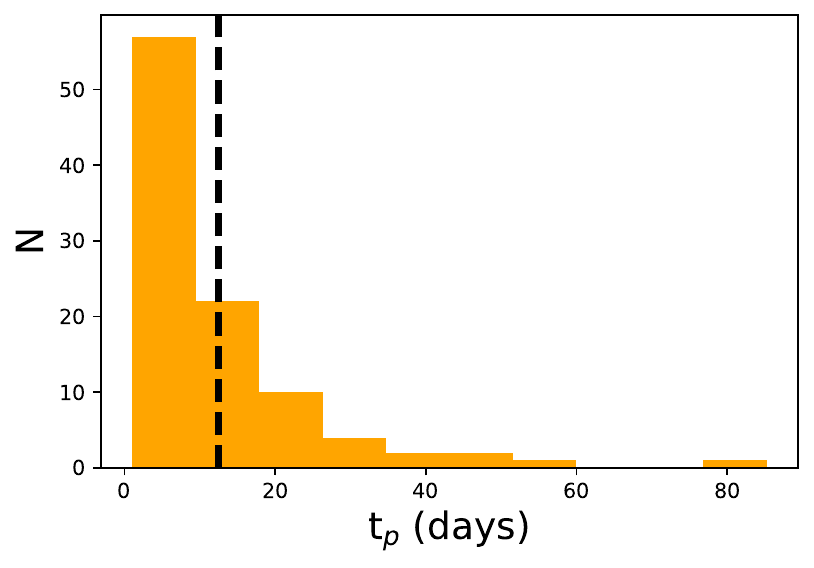}
\includegraphics[width=6cm]{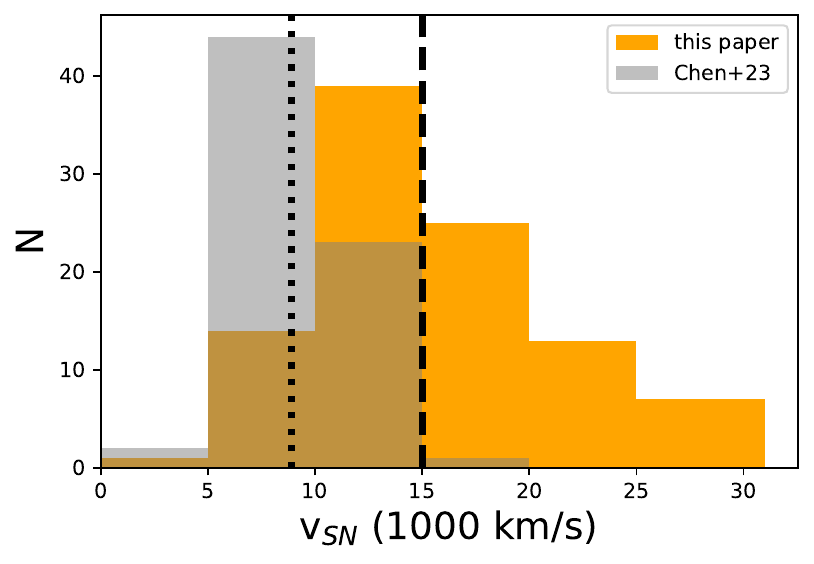}
\includegraphics[width=6cm]{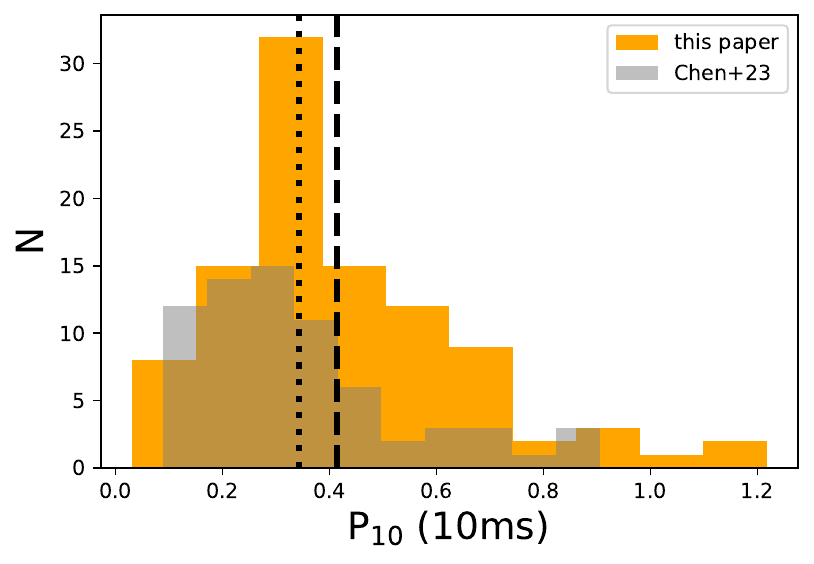}
\includegraphics[width=6cm]{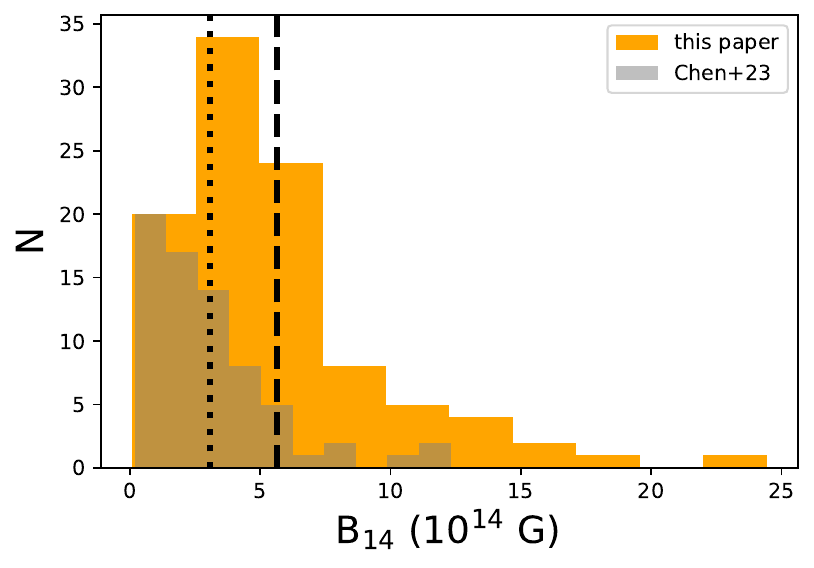}
\includegraphics[width=6cm]{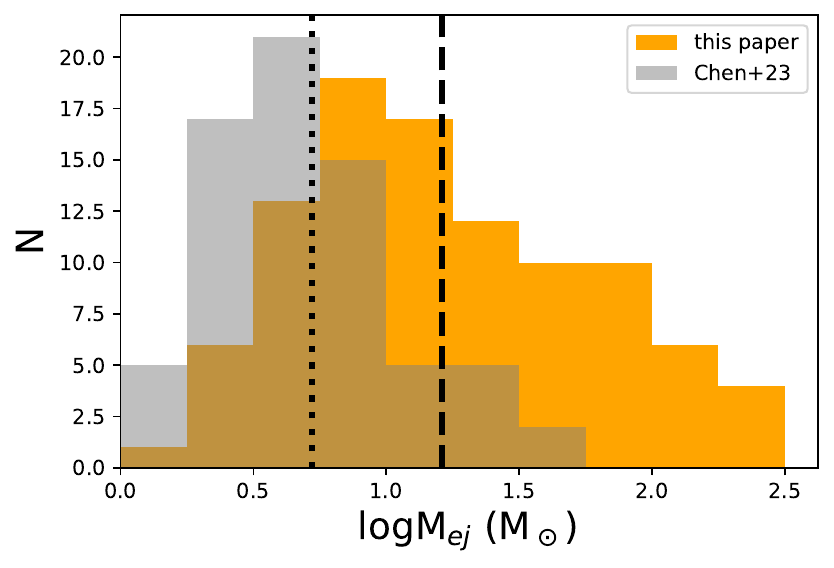}
\caption{Histograms, showing the distribution of the fitted and inferred parameters of the magnetar model. Orange color denotes to our results, together with the mean value of each parameter shown with dashed vertical lines, while the values obtained by \citet{chen23b} are plotted with grey, and their mean is represented with dotted vertical lines.}
\label{fig:hist_mag}
\end{figure*}

The magnetar model of Minim has 6 free parameters that are optimized by the code when fitting the observed light curves
($t_{\rm ini}$, $R_0$, $E_p$, $t_{\rm diff}$, $t_p$ and $v_{\rm SN}$), from which the spin-down period, the magnetic field and the ejected mass can be calculated via Eq. \ref{eq:p10b14} and \ref{eq:mej} in order to compare them to the results of \citet{chen23b}. It is important to note here that MOSFiT fits $B_\perp$, the component of the magnetic ﬁeld perpendicular to the spin axis, while Minim fits a different $B_{14}$ defined by \citet{2010ApJ...717..245K}, where the angle between the magnetic axis and the line of sight is 45$^\circ$. Therefore the following conversion was applied to make the B-values comparable with each other: $B_\perp~=~B_{B14}/2.5$ \citep[see][for the detailed calculations of the conversion]{2017ApJ...850...55N}.

In Figure \ref{fig:hist_mag}, histograms showing the distribution of the fitted/interred parameters referring to the magnetar model are plotted. Orange color denotes to the results of this paper, while grey color shows the values obtained from \citet{chen23b}.

It can be seen in this figure, as well as in Figure \ref{fig:irodalom} that the $P_{10}$ and $B_{14}$ parameters inferred in this paper are consistent with those of \citet{chen23b}, while the ejected masses obtained here are significantly larger compared to \citet{chen23b}. 

One of the factors that may cause the differences between the ejecta masses of the two studies can be found in the $v_{\rm SN}$ values, which are averagely larger in this paper ($v_{\rm SN, avg}~=~15010$ km s$^{-1}$) compared to \citet{chen23a} ($v_{\rm SN, avg}~=~8904$ km s$^{-1}$). 
The velocity ratio between the average velocities of the overlapping SNe that are present in both the sample analyzed here and the sample studied by \citet{chen23b} were calculated to be $v_{\rm SN, this paper} / v_{\rm SN, Chen}~=~1.76$.

The lower velocities of \citet{chen23a} originate from spectroscopic measurement, similarly to this study, however, they used the Fe II and O II lines in the spectra to calculate the  $v_{\rm SN}$, while here, cross-correlation was applied. When spectroscopy was not available, \citet{chen23a} used bounds between 3000$-$25000 km s$^{-1}$, while in this study it was set to 8000$-$30000 km s$^{-1}$. It can be seen that the  different method to estimate the velocities resulted in significantly larger $v_{\rm SN}$ in the present paper. The velocity values used here do not contradict to the physics of SLSNe-I, that usually show larger velocities compared to normal SNe. 
It is important to note that the extinction due to the host galaxy, which contributes to the observed light curves, was neglected in this paper, while it was taken into account by MOSFiT in the study of \citet{chen23b,chen23a}. The extinction changes only the amplitude of the light curve and does not affect the shape of the light curve. By modeling bolometric light curves instead of multi-color data, the effects of host extinction can be somewhat reduced. With neglecting it, the ejecta masses from the modeling become a little lower. Therefore if one gets unrealistically low ejecta mass estimates, it is likely caused by neglecting the extinction. However, this is not the case in this study, where the inferred ejecta masses tend to be larger, compared to other works. Furthermore, the host extinction as a fitting model parameter decreases the degree of freedom, thus, increases the uncertainty of the other fitted parameters. 

Another cause of the difference between the ejected masses inferred in these two studies can be the usage of different priors in the models. While Minim works with flat prior distributions, the prior can be modified for each free parameter in MOSFiT. \citet{chen23a} also used a flat distribution to the velocities, however, for other parameters, they used priors similar to \citet{2017ApJ...850...55N}, where flat, Gaussian and log-flat distributions were applied as well. From the other studies used for the comparison of the magnetar models,  \citet{2021ApJ...921..180H} and \citet{2022ApJ...937...13H}  presented the applied priors, where flat (for B, P and M), log-flat (e.g. for $\kappa_\gamma$) and Gaussian (for $v_{SN}$) distributions were listed as well. 
The usage of non-flat priors injects a trend into the system artificially, thus, a lot of possible solutions are excluded from the parameter space.
It is difficult to decide which mass estimates are closer to the real, physical values, however, our average $M_{\rm ej}$ of $\sim 20 M_\odot$ reinforces the assumption that SLSNe-I originate from the explosion of initially very massive stars.

\begin{figure*}[h!]
\centering
\includegraphics[width=8cm]{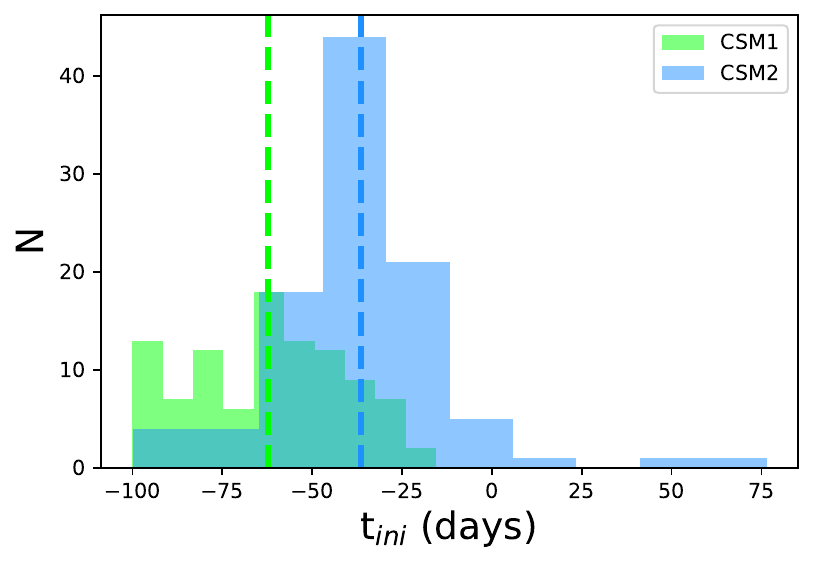}
\includegraphics[width=8cm]{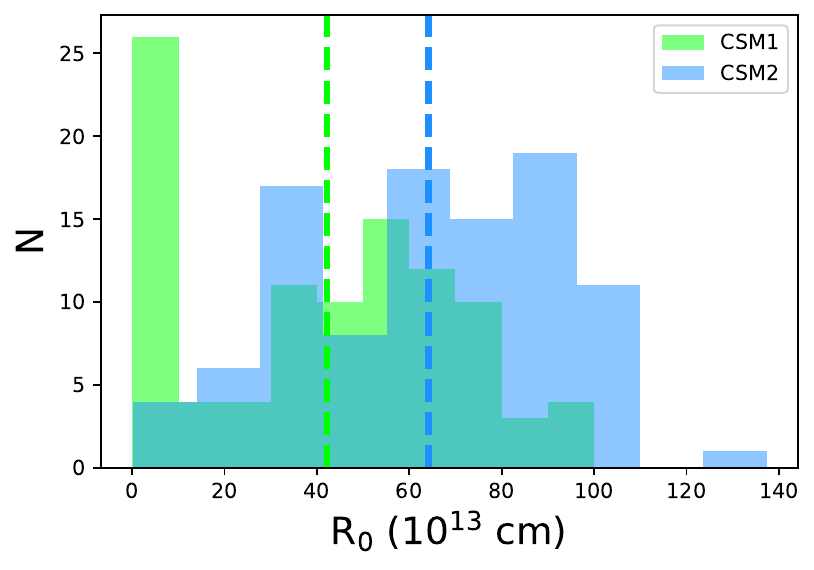}
\includegraphics[width=8cm]{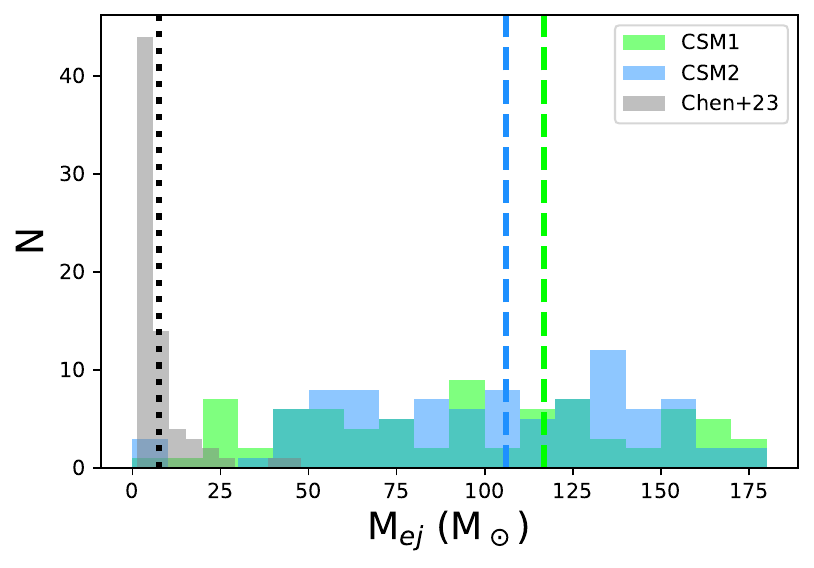}
\includegraphics[width=8cm]{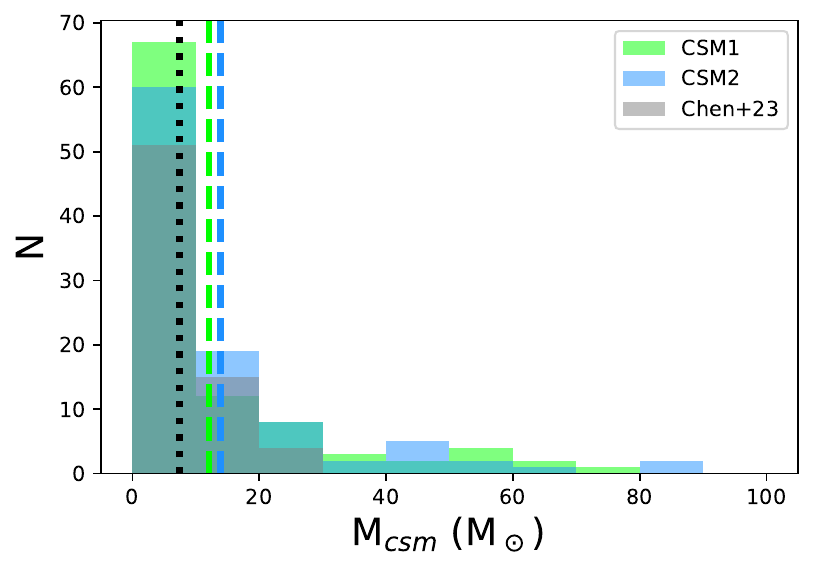}
\includegraphics[width=8cm]{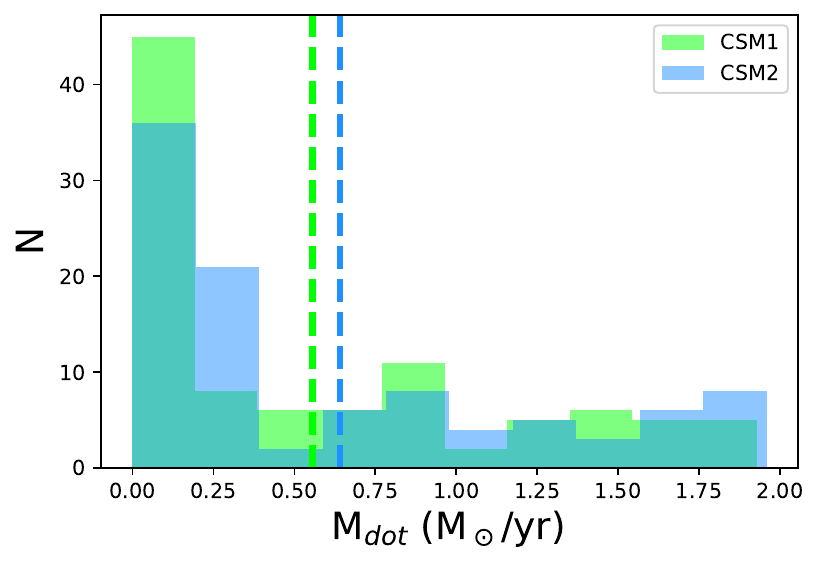}
\includegraphics[width=8cm]{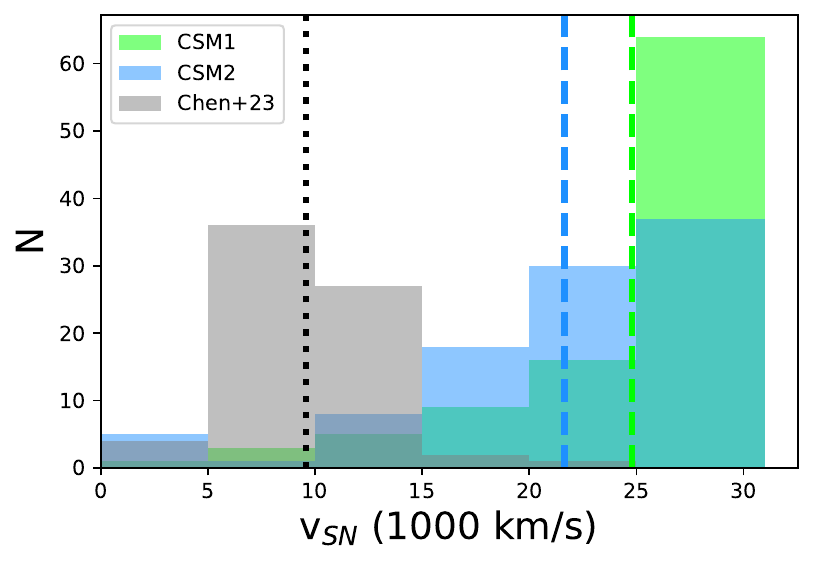}
\caption{Histograms, showing the distribution of the fitted and inferred parameters of the CSM (green) and the CSM 2 (blue) models, compared to the results of \citet{chen23b} (grey). The dashed and the dotted lines have the same meaning as in Figure \ref{fig:hist_mag}.}
\label{fig:hist_hibrid}
\end{figure*}

\begin{figure}[h!]
\centering
\includegraphics[width=8cm]{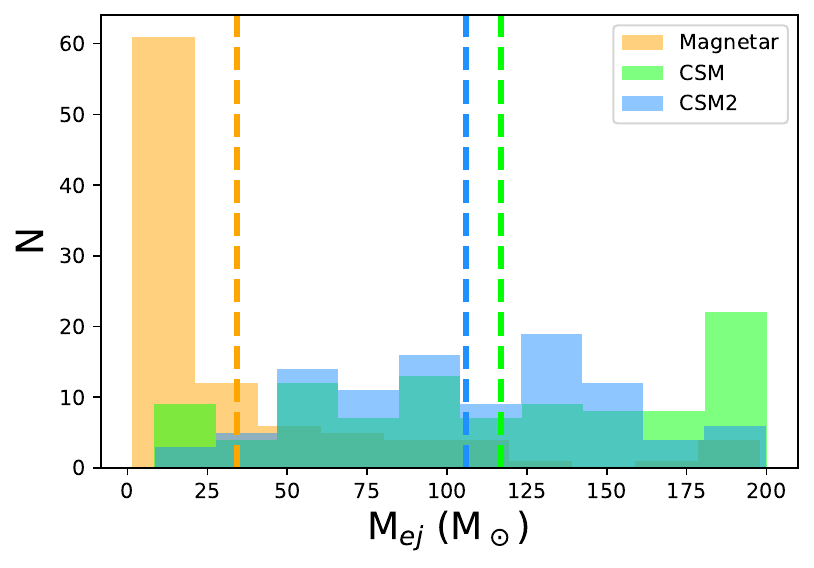}
\caption{The distribution of the ejecta masses from the magnetar (orange), the CSM (green) and the CSM 2 (blue) models. The mean values of each model are shown with dashed vertical lines.}
\label{fig:hist_mejossz}
\end{figure}

\subsubsection{The comparison of the CSM models}

Secondly, the CSM modeling of \citet{chen23b, chen23a} were compared to our results. The CSM models (both the constant density and the steady wind model) of Minim have the following fitted parameters: $t_{\rm ini}$, $R_0$, $M_{\rm ej}$, $M_{\rm CSM}$, $M_{\rm dot}$ and $v_{\rm SN}$. Here, the effects of radioactive decay were turned out, therefore $M_{\rm Ni}$ was equaled 0 in order to decrease the degree of degeneracy. On the other hand, \citet{chen23b}
modeled their light curves with the collective presence of both radioactive decay and CSM-interaction. The fitted parameters of MOSFiT are $M_{\rm CSM}$, $M_{\rm ej}$, $M_{\rm Ni}$ and $E_{\rm k}$. Here the following question arises: are the results of \citet{chen23b,chen23a} comparable to the results of this paper? With the Ni-heating turned on, it is possible that instead of the interaction, the radioactive decay dominates the light curve, and therefore the resulting CSM mass and ejecta mass becomes naturally lower compared to the models that assume purely interaction, without radioactive decay. 

In order to answer this question, a test was made: the CSM+Ni model of Minim was fitted to the light curve of 5 well observed SLSNe-I, which are present in the sample of \citet{chen23b,chen23a} as well, and the resulting parameters were compared to each other. The 5 SLSNe-I are SN2018bym, SN2019eot, SN2019neq, SN2019nhs and SN2020exj. To the quasi-bolometric light curves of these objects, the constant density model CSM (CSM1) was fitted using Ni-masses in the range between 0 and 10 $M_\odot$. The resulting model light curves are plotted with red lines in Figure \ref{fig:bol} in the Appendix, while Table \ref{tab:ni_par} in the Appendix collects the best-fit values for all fitted parameters.

\begin{figure}[h!]
\centering
\includegraphics[width=8cm]{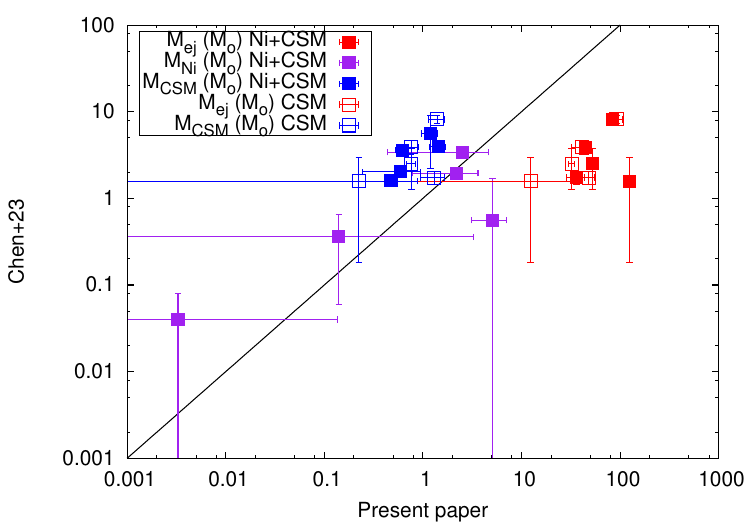}
\caption{Comparison of the best-fit model parameters of 5 well-observed SLSNe-I between \citet{chen23a} and this paper. Filled squares denote to the Ni+CSM model of Minim, while with empty squares, the constant velocity CSM model of Minim is shown. The best-fit ejecta masses (red), nickel masses (purple) and CSM masses (blue) are compared to each other. In conclusion, the nickel masses and CSM masses obtained by \citet{chen23a} are in agreement with our results, while this paper shows an order of magnitude larger ejecta masses (in case of both the CSM+Ni and the CSM models), then the MOSFiT modeling of \citet{chen23a}. }
\label{fig:mejcompare}
\end{figure}

Figure \ref{fig:mejcompare} makes a comparison between the fitted $M_{\rm ej}$, $M_{\rm Ni}$ and $M_{\rm CSM}$ parameters of the best-fit CSM+Ni model of \citet{chen23a}, the best-fit CSM+Ni model and the best-fit CSM model of the present paper. The conclusions are the following:

\begin{itemize}

    \item $M_{\rm Ni}$ ($M_\odot$; purple filled squares in Figure \ref{fig:mejcompare}): it can be seen that the Ni-masses in case of both the models of \citet{chen23a} and the models presented here are in agreement within errors. Although the upper limit of $M_{\rm Ni}$ was 10 $M_\odot$ in this paper, the resulting nickel masses show a range between 0.01 and 2.54 $M_\odot$, similarly to \citet{chen23a} for the 5 comparison objects. In the fits (see the red curves in Figure \ref{fig:bol} in the Appendix), it can be seen that when the Ni is turned on, it affects mostly the rising part of the light curves.
    \item $M_{\rm ej}$ ($M_\odot$; red squares in Figure \ref{fig:mejcompare}): the ejecta masses estimated by Minim exceed by an order of magnitude the values derived by \citet{chen23a}. For the 5 selected SLSNe-I, the MOSFiT modeling resulted in masses between 1 and 10 $M_\odot$, while the CSM+Ni model of Minim gave masses between $\sim$36 and $\sim$124 $M_\odot$. Filled quares in Figure \ref{fig:mejcompare} refer to the CSM+Ni model of Minim, while the empty squares compare the purely CSM fit of Minim to the results of \citet{chen23a}. It can be seen that the ejecta masses inferred by the two Minim models (i.e. the Ni+CSM model and the CSM model) agree with each other, and both are an order of magnitude larger compared to the MOSFiT modeling of \citet{chen23a}. 
    \item$M_{\rm CSM}$ ($M_\odot$; blue squares in Figure \ref{fig:mejcompare}): the mass of the CSM shell was estimated to be in between 0.48 and 1.43 $M_\odot$ by the Ni+CSM model of Minim for the 5 well-observed test-objects. These values are consistent with both the pure CSM model of Minim, and the models of \citet{chen23a}, although in \citet{chen23a}, the CSM mass is a little bit larger, compared to the results of this paper.  
\end{itemize}

Globally, it can be stated that turning the Ni-heating on in the model of Minim has little impact on the ejecta mass estimates, therefore it makes sense to compare the parameters from the pure CSM models with the hibrid model of \citet{chen23a}. (According to \citet{chen23a} some light-curves were modeled using Ni-masses that exceed 10-20 $M_\odot$, which may be considered as unrealistically high.)
Initially, the Ni-masses were equaled with 0 in the models to decrease the number of free parameters, and such the degree of degeneracy and the run-time of the models. After this test with the 5 well-observed objects it is concluded that modeling every SLSNe-I in the sample is not necessary for the comparison of our results to those of \citet{chen23b}. In addition, setting $M_{\rm Ni}>0$ in the models gives an extra heating source, which is not in direct connection with the ejected mass. However, in a model that assumes purely CSM interaction, the overall luminosity comes from the collision between the ejecta and the CSM, which is directly related to the ejected mass. In the CSM models, the excess luminosity requires more mass on the contrary to the CSM+Ni models, where the Ni gives an extra heating to the system.

Figure \ref{fig:hist_hibrid} shows the results of the modeling in histograms, where green color denotes to the CSM model, blue codes the CSM 2 model and the best-fit parameter values obtained by \citet{chen23b} are plotted with grey.

These plots show that the CSM 1 and the CSM 2 model of Minim gives similar results. (The mean values of each parameter can be found in Table \ref{tab:csm_par}.) The modeling of \citet{chen23b}
resulted in a mean $M_{\rm CSM}$ of 4.67 $M_\odot$ with a similar distribution to the CSM and the CSM 2 models, which have averages of 12.06 and 13.90 $M_\odot$, respectively. However, the ejecta mass values of \citet{chen23b} have a mean of 11.92 $M_\odot$, which is significantly smaller than the the ejecta masses estimated by the CSM and the CSM 2 models: 116.82 and 105.99 $M_\odot$, respectively.

It can be seen that the results of \citet{chen23b,chen23a} show smaller parameter ranges and values in general. Although MOSFiT uses the same equations in the models as \citet{2012ApJ...746..121C}, there are differences between the results of \citet{chen23b} and this study. One of the reasons behind the discrepancy between the ejecta masses can be found in the value of $\kappa$, which was fitted in a range of 0.05-0.34 in the CSM models of \citet{chen23b}, while it was fixed to 0.2 in this paper. The ejected mass values are very sensitive to $\kappa$: a change of 0.1 in $\kappa$ results in a change in $M_{\rm ej}$ by a factor of 2. Second, $M_{\rm ej}$ depends strongly on the parameter $v_{\rm SN}$, which is estimated to be larger in this paper, compared to \citet{chen23a}, similarly to the case of the magnetar modeling. 

Another factor that plays a large role in the resulting ejecta masses of the models is the efficiency of converting kinetic energy into radiation. According to \citet{2018ApJ...869..166V}, the efficiency used in MOSFiT is 0.5, while Minim assumes that the 100\% of the kinetic energy is converted into radiation. This makes the discrepancy between the ejecta masses shown in this paper and in the $M_{\rm ej}$ values presented by e.g. \citet{chen23a} even worse.

Figure \ref{fig:hist_mejossz} shows the distribution of ejected masses calculated from the magnetar, the CSM and the CSM 2 model of Minim. It can be seen that the magnetar model gives significantly lower values for $M_{\rm ej}$, then the CSM model, but with an average value of $\sim 25 M_\odot$, it assumes a large mass progenitor star. According to the CSM models, a lot of SLSNe originate from the explosion of the most massive stars, which are initially more than 100 $M_\odot$, and may eject 20 $M_\odot$ of circumstellar material before explosion. This is consistent with \citet{2017hsn..book..403S}, who summarized the properties of interacting SNe, and showed that interacting superluminous supernovae may eject from the lowest to the highest CSM masses, and most likely originate from the explosion of the largest stars. It is important to note that 
an ejecta mass larger than 100 $M_\odot$ means that the initial mass of the star has to be even larger than that value, or the explosion happened in a merger double system. However, such stars/systems are very rare or unlikely, which shows the main caveat of using a pure CSM model to describe the observed luminosity. Thus, the ejecta mass estimates obtained using the CSM model of Minim may overestimate the real, physical ejected masses, and give upper limits instead of reliable estimates, even though they show that the light curves of SLSNe can be fitted well using the largest masses possible. Therefore these results have to be treated cautiously, since these models have a lots of free parameters, and a minor change in one of the them may result in much higher or lower ejecta masses.

\section{Summary}\label{sec:sum}

We present the bolometric light curve modeling of 98 superluminous supernovae that exploded before 2024 to compare our results with those obtained by \citet{chen23b,chen23a} using MOSFiT and to other earlier studies. To be consistent with \citet{chen23b,chen23a}, the quasi-bolometric light curves were constructed from the ZTF g- and r-band data using the formula of \citet{chen23b}, which relates the $(g-r)$ color index to the bolometric luminosity. They were then modeled using the magnetar and CSM input of the Minim code, which is based on the radiation diffusion model of \citet{1980ApJ...237..541A,1982ApJ...253..785A}. 

The magnetar model of Minim has 6 fitted parameters, which are that of the explosion with respect to the date of the peak brightness, the initial radius of the progenitor, the magnetar rotation energy, the diffusion time scale, the spin-down time scale of the magnetar and the ejecta velocity. From these, the  mean value of the initial spin period of the magnetar and its magnetic field were calculated as $P~=~4.1 \pm 0.20$ ms and $B~=~5.65 \pm 0.43 \cdot 10^{14}$ G. The ejected masses were derived from the formula of \citet{1980ApJ...237..541A}, where $M_{\rm ej}$ is proportional to the ejecta velocity. The latter parameter was constrained from spectroscopy, and in case of the SLSNe-I that had available spectra in WiseRep, the ejecta velocity was fixed to the estimated value. While for the objects without inferred velocities, it was fitted in between 8000 and 30000 km s$^{-1}$ by Minim. The final ejected mass was estimated to be 34.26 $\pm$ 4.67 $M_\odot$.

While the obtained P and the B parameters were found to be consistent with previous studies in the literature \citep[see][]{2015MNRAS.452.3869N, 2017ApJ...840...12Y, 2017MNRAS.470.3566C, 2018ApJ...860..100D, 2018ApJ...864...45M, 2018MNRAS.479.4984C, 2019ApJ...874...68C, 2019MNRAS.487.2215A,2021ApJ...921..180H, 2022ApJ...937...13H, chen23b, chen23a}, our ejected masses are globally higher compared to the earlier results: in most studies the magnetar modeling of the light curves of a large sample of SLSNe-I resulted in mean ejected masses of $\sim 4 - 10 M_\odot$. Besides the used $\kappa$ values, one of the main reasons behind the discrepancy of the ejecta masses may be found in the different ejecta velocities: the MOSFiT modeling presented by \citet{chen23b}, as well as other studies gave considerably lower $v_{\rm SN}$ values compared to the usually high ejecta velocities of SLSNe-I \citep[see e.g.][]{2018ApJ...854..175I,2023ApJ...954...44K}.

After applying the magnetar model, the CSM interaction input of Minim was used  to reproduce the observed light curves. In this case, the fitted parameters were the time between the explosion and the maximum,  the radius of the progenitor, the ejected mass, the pre-SN wind mass loss rate and the ejecta velocity.  Radioactive nickel decay was turned off to reduce the degrees of freedom. Two different models were applied to this explosion scenario: the constant density ($s~=~0$) and the steady wind ($s~=~2$) models (named CSM and CSM2, respectively). It is concluded that the value of the ejecta mass is significantly larger compared to the $M_{\rm ej}$ values from the magnetar models: we obtained  116.82 $\pm$ 5.97 for the constant density model and 105.99 $\pm$ 4.50 for the steady wind model. Both these values exceed the mean ejecta mass estimated by \citet{chen23b} ($11.92 M_\odot$), but are consistent with \citet{2013AAS...22123305C}, who used the Minim code to model the bolometric light-curves of a small sample of SLSNe-I. Our CSM mass estimates also exceed the estimates of \citet{chen23a}: we obtained 12.06 $\pm$ 1.74  $M_\odot$ and 13.90 $\pm$ 1.67 $M_\odot$ for the CSM and the CSM2 models, respectively, while \citet{chen23a} had an average CSM mass of 7.44 $M_\odot$. The cause of the inconsistency in the ejected masses may lie in the different ejecta velocity values, and the different input bounds, proxies, and model set-ups. The $\kappa$ parameter also plays a significant role in the resulting masses. While \citet{chen23b} fitted the $\kappa$ between 0.05 and 0.34, it was set to 0.2 in this paper, which can lead to large differences. Therefore both results have to be treated cautiously.

By examining the reduced $\chi^2$ values of the individual objects, it was found that 14 SLSNe-I favor the magnetar model, 39 objects prefer the CSM model, and the light curve of 45 SLSNe-I can be fitted equally well with both types of models.

It is important to note that our modeling does not change the overall picture of SLSNe-I: they are most likely the deaths of the most massive stars, which eject up to tens of solar masses in their explosions. Although the ejecta masses calculated here are considerably higher than in previous studies, it is possible that SLSNe-I eject 20 or even 100 $M_\odot$ when they explode. Since the CSM and the magnetar model could fit the quasi-bolometric light curves equally well on average, the distinction between the two explosion scenarios and the setting up of the astrophysical big picture on the brightest stellar explosions (including the possibility of pair-instability) require further analysis of large samples of SLSNe-I, which is left to future studies.

\section{Data availability}\label{sec:available}

Supplementary data to the paper, such as plots of the bolometric light-curve  modeling and velocity calculation process, as well as tables collecting the basic data and the model parameters of the studied SLSNe-I can be found in the following website: \url{ https://doi.org/10.5281/zenodo.14591928}.

\acknowledgments

This project is supported by the NKFIH/OTKA FK-134432 and the NKFIH/OTKA K-142534 grant of the National Research, Development and Innovation (NRDI) Office of Hungary. We are indebted to the anonymous referee, who helped with constructive criticism and suggestions to improve the quality of the paper.

\newpage

\bibliography{main.bib} 
\bibliographystyle{aasjournal}

\appendix

\setcounter{table}{0}
\renewcommand{\thetable}{A\arabic{table}}

\setcounter{figure}{0}
\renewcommand{\thefigure}{A\arabic{figure}}

\begin{table*}
\caption{SLSNe removed from our sample. }
\label{tab:removed}
\begin{center}

\begin{tabular}{ll}
\hline
\hline
Reason for exclusion [number] & Object \\ 
\hline 
 \multicolumn{2}{c} { \bf From the SLSN-I sample of WiseRep that have ZTF name [78]} \\
\hline
 Unrealistic/messy LC[18] & SN2018avk, SN2018bgv, SN2018ibb, SN2019J, SN2019cca, \\ & SN2019pvs, SN2020ful, SN2020onb,  SN2020rmv, \\
  & SN2020tcw, SN2021lji, SN2021lwz, SN2021tkx, SN2021hpc, \\ & SN2021hpx, SN2023hoz, SN2023vco, SN2023aafk \\
 \hline 
 Classified as non-SLSN-I [4] &  SN2019aanx (SLSN-II), SN2019pud (SLSN-II), SN2020jhm (SLSN-II), SN2022csn (TDE) \\
\hline
 \multicolumn{2}{c}{\bf{From the sample of \citet{chen23b} [78]} }\\
\hline
Classified as non-SLSN-I [4] & SN2018gkz (Ia), SN2019gfm (BL-Ic), SN2019hge (IIb), SN2019unb (II) \\
\hline
Other reasons [14] &   SN2018avk, SN2018bgv, SN2018lzw, SN2018lzx, SN2018hpq, SN2018hti, \\ &  SN2019J, SN2019cca, SN2019cwu,  SN2019fiy, SN2019onb, SN2019otl, \\ & SN2019rmv, SN2019vvc, SN2019aamu\\
\hline

\end{tabular}

\end{center}
\end{table*}

\begin{figure*}[h!]
\centering
\includegraphics[width=16cm]{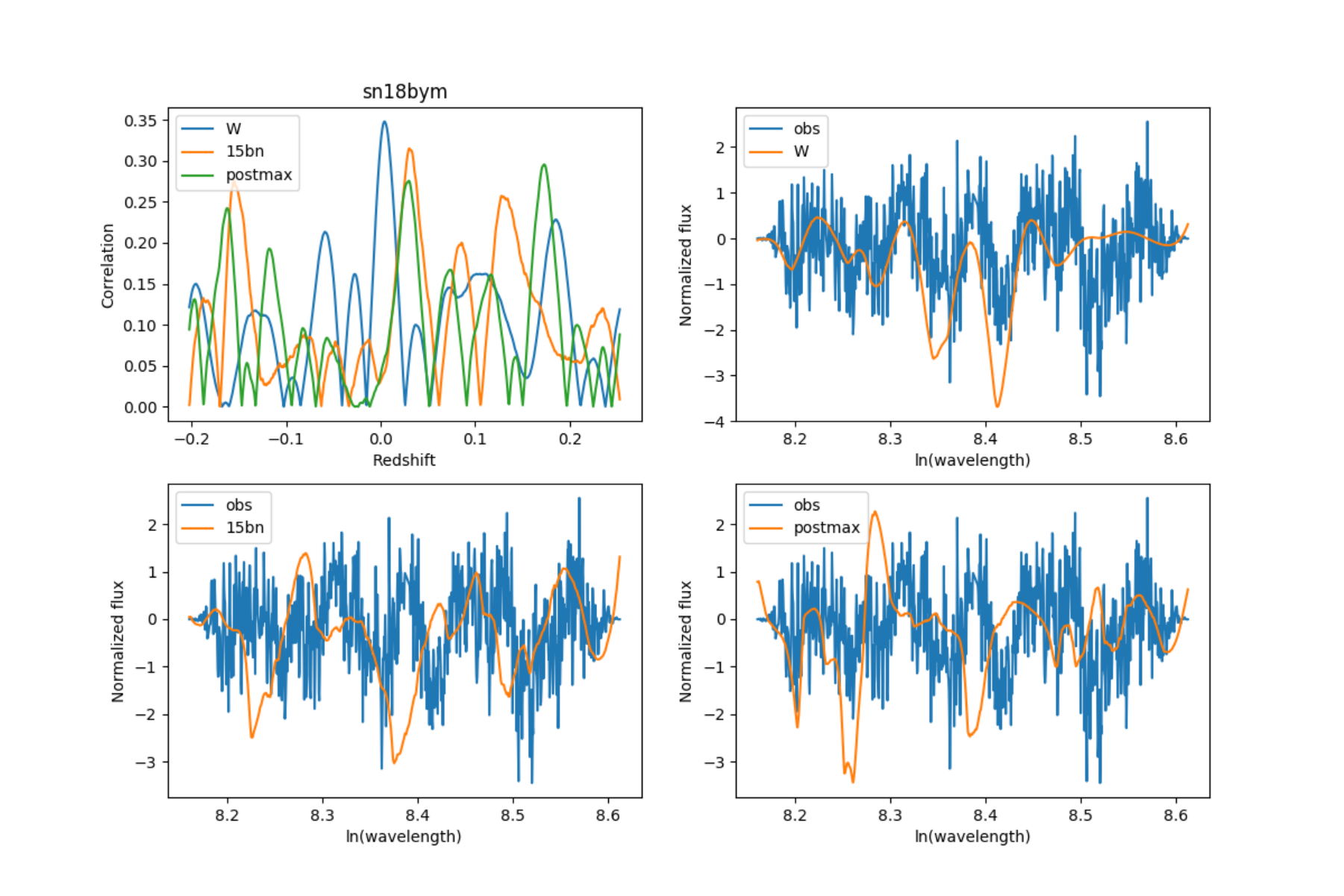}
\caption{ The cross-correlation of the available spectra with the templates published in \citet{2021ApJ...909...24K}. The upper left panel shows the redshift vs. correlation with different colors for each templates. The upper right, the bottom left and the bottom right panels plot together the observed spectrum of a particular object (blue) with the Type W, the Type 15bn and the post-maximum templates (orange), respectively. In this example, the best match of the observed spectrum of SN2018bym is the Type W model in the top right panel, therefore the velocity of SN2018bym is estimated from the redshift of the 10000 km s$^{-1}$ version of the that model. Similar plots for each object in the sample are collected in the online supplementary material: \url{ https://doi.org/10.5281/zenodo.14591928}.} 

\label{fig:fxcor}
\end{figure*}

\begin{figure*}[h!]
\centering
\includegraphics[width=8cm]{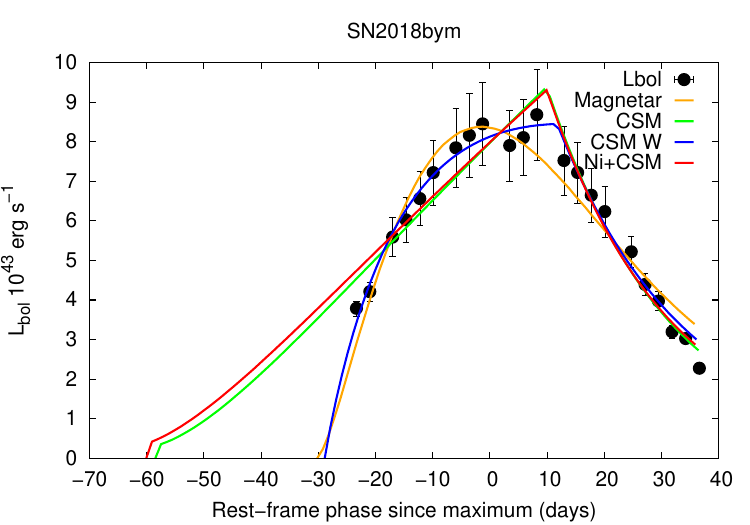}
\includegraphics[width=8cm]{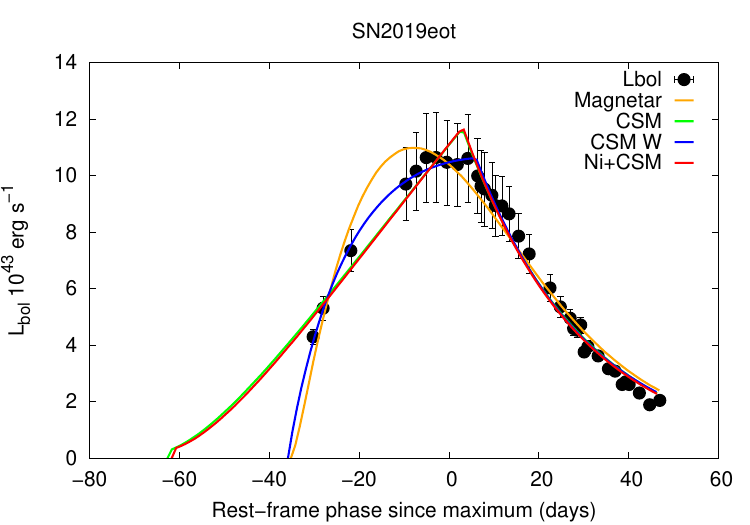}
\includegraphics[width=8cm]{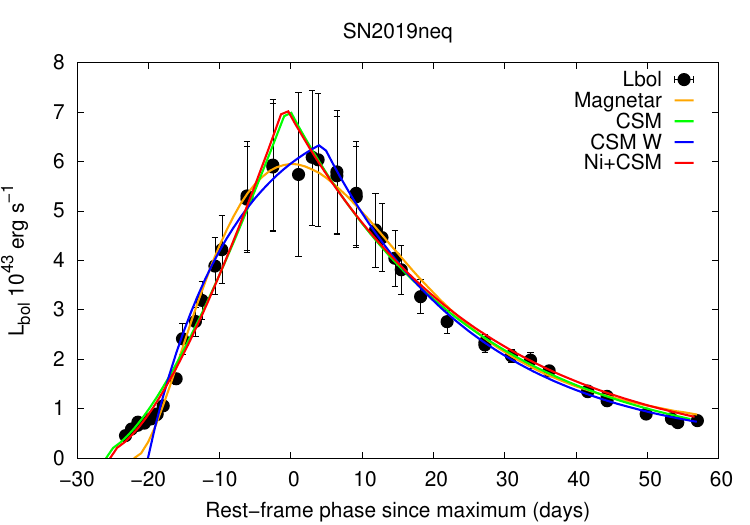}
\includegraphics[width=8cm]{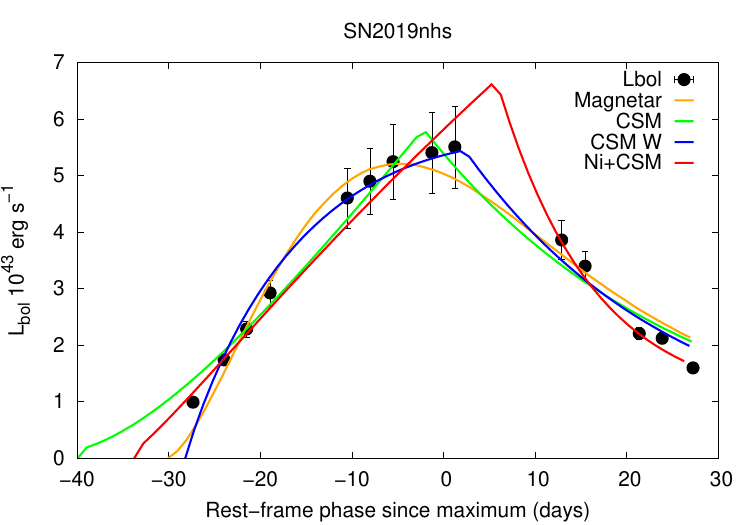}
\includegraphics[width=8cm]{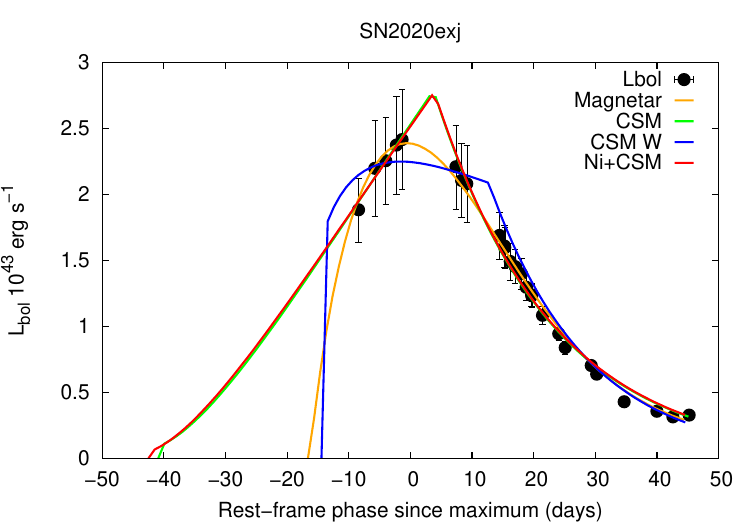}
\caption{The quasi-bolometric light curves of the studied SLSNe-I (black dots) plotted together with their best-fit magnetar (orange), CSM (green), CSM 2 (blue) and Ni+CSM (red) models obtained using Minim. Similar plots for all studied SLSNe-I are collected in the online supplementary material: \url{ https://doi.org/10.5281/zenodo.14591928}.} 
\label{fig:bol}
\end{figure*}

\begin{table*}[h!]
\caption{The best-fit parameters of the magnetar model.  The full table is available in the online supplementary material: \url{ https://doi.org/10.5281/zenodo.14591928}.}
\label{tab:magnetar_par}
\begin{center}
\scriptsize
\begin{tabular}{lccccccc}
\hline
\hline
SLSN & $t_{\rm ini}$  & $R_0$&$E_{\rm p}$ & $t_{\rm diff}$ & $t_{\rm p}$ & $v_{\rm SN}$ & $\chi^2$ \\
     &  (days) & $10^{13}$cm &($10^{51}$ erg) & (days) &  (days) & (1000 km s$^{-1}$) & \\            
\hline
SN2018bym & -23.06 (0.33) & 1.00 (0.01) & 1.08 (0.03) & 24.02 (0.67) & 11.01 (1.12) & 11.02 (1.52) & 4.7311 \\
SN2018don & -48.10 (1.86) & 6.49 (2.32) & 0.14 (0.02) & 58.87 (2.98) & 6.24 (1.24) & 24.28 (0.01) & 2.3738 \\
SN2018ffs & -32.11 (1.69) & 5.41 (0.75) & 0.60 (0.15) & 41.97 (6.28) & 7.58 (1.73) & 12.26 (0.01) & 0.9581 \\
\bf{...} & & & & & & & \\
\hline
\end{tabular}
\end{center}
\end{table*}

\begin{table*}[h!]
\caption{The P, B and $M_{\rm ej}$ parameters of the magnetar model derived from the fitted parameter values that are shown in Table \ref{tab:magnetar_par}.   The full table is available in the online supplementary material: \url{ https://doi.org/10.5281/zenodo.14591928}.}
\label{tab:magnetar_par2}
\begin{center}
\scriptsize
\begin{tabular}{lccc}
\hline
\hline
SLSN & P$_{10}$ & B$_{14}$ & $M_{\rm ej}$ \\
     & (10 ms) & ($10^{14}$ G) & ($M_\odot$) \\
\hline
SN2018bym & 0.031 (4.433) & 0.091 (4.912) & 1.53 (0.01) \\
SN2018don & 1.217 (0.018) & 16.649 (0.093) & 32.51 (2.64) \\
SN2018ffs & 0.578 (0.154) & 7.183 (0.098) & 8.34 (3.68) \\
\bf{...} & & & \\
\hline
\end{tabular}
\end{center}
\end{table*}

\begin{table*}[h!]
\caption{The best-fit parameters of the CSM model.  The full table is available in the online supplementary material: \url{ https://doi.org/10.5281/zenodo.14591928}.}
\label{tab:csm_par}
\begin{center}
\scriptsize
\begin{tabular}{lccccccc}
\hline
\hline
SLSN & $t_{\rm ini}$ & $R_0$ & $M_{\rm ej}$ & $M_{\rm CSM}$ & $\dot{M}$ & $v_{\rm SN}$ & $\chi^2$ \\
     &  (days)       &$10^{13}$cm & ($M_\odot$)  & ($M_\odot$)  & ($M_\odot$/yr)  & (1000 km s$^{-1}$) & \\ 
\hline
SN2018bym & -58.51 (3.86) & 66.91 (14.83) & 92.03 (13.77) & 1.39 (0.26) & 0.802 (0.240) & 27.32 (1.51) & 2.7742 \\
SN2018don & -66.04 (16.96) & 58.67 (30.60) & 24.11 (50.70) & 11.35 (39.31) & 1.437 (0.437) & 18.13 (3.61) & 0.2816 \\
SN2018ffs & -41.00 (8.22) & 21.38 (14.69) & 8.38 (28.04) & 3.16 (47.71) & 0.005 (0.219) & 29.56 (1.15) & 1.8073 \\
\bf{...} & & & & & & & \\
\hline
\end{tabular}
\end{center}
\end{table*}

\begin{table*}[h!]
\caption{The best-fit parameters of the CSM2 model.  The full table is available in the online supplementary material: \url{ https://doi.org/10.5281/zenodo.14591928}.}
\label{tab:csm2_par}
\begin{center}
\scriptsize
\begin{tabular}{lccccccc}
\hline
\hline
SLSN & $t_{\rm ini}$ & $R_0$&$M_{\rm ej}$ & $M_{\rm CSM}$ & $\dot{M}$ & $v_{\rm SN}$ & $\chi^2$ \\
     &  (days)       &$10^{13}$cm& ($M_\odot$)  & ($M_\odot$)  & ($M_\odot$/yr)  & (1000 km s$^{-1}$) & \\ 
\hline
SN2018bym & -28.88 (0.84) & 81.41 (15.30) & 135.00 (28.25) & 4.19 (0.34) & 0.103 (0.012) & 26.30 (0.99) & 1.3981 \\
SN2018don & -50.28 (25.30) & 66.38 (35.08) & 35.35 (51.45) & 14.64 (31.36) & 1.898 (1.072) & 9.19 (0.99) & 0.286 \\
SN2018ffs & -33.91 (1.49) & 83.36 (10.59) & 94.12 (29.30) & 7.28 (3.87) & 0.337 (0.277) & 17.97 (0.80) & 0.8954 \\
\bf{...} & & & & & & & \\
\hline
\end{tabular}
\end{center}
\end{table*}

\begin{table*}[h!]
\caption{The best-fit parameters of the CSM+Ni model.}
\label{tab:ni_par}
\begin{center}
\scriptsize
\begin{tabular}{lcccccccc}
\hline
\hline
SLSN & $t_{\rm ini}$ & $R_0$&$M_{\rm ej}$ & $M_{\rm CSM}$ & $\dot{M}$ & $M_{\rm Ni}$ &$v_{\rm SN}$ & $\chi^2$ \\
     &  (days)       &$10^{13}$cm& ($M_\odot$)  & ($M_\odot$)  & ($M_\odot$/yr) & ($M_\odot$) & (1000 km s$^{-1}$) & \\ 
\hline
SN2018bym & -60.08 (3.77) & 70.32 (12.03) & 84.54 (9.67) & 1.43 (0.25) & 0.698 (0.172) & 2.54 (2.10) & 28.09 (1.09) & 2.9023 \\
SN2019eot & -61.74 (2.82) & 46.51 (8.59) & 124.10 (13.23) & 1.19 (0.21) & 0.659 (0.166) & 2.21 (1.42) & 26.69 (1.44) & 0.9225 \\
SN2019neq & -25.38 (0.98) & 23.84 (4.89) & 44.39 (5.88) & 0.62 (0.09) & 0.959 (0.479) & 0.14 (3.11) & 28.35 (1.96) & 1.0322 \\
SN2019nhs & -33.77 (3.88) & 60.70 (14.90) & 36.22 (7.38) & 0.59 (0.35) & 1.681 (0.545) & 5.11 (1.97) & 25.38 (2.18) & 0.9224 \\
SN2020exj & -42.47 (2.42) & 26.31 (2.21) & 52.58 (2.03) & 0.48 (0.04) & 0.946 (0.073) & 0.01 (0.13) & 17.00 (0.25) & 4.1465 \\
\hline
\end{tabular}
\end{center}
\end{table*}

\end{document}